\documentclass[sigconf]{aamas}

\usepackage{balance} 

\usepackage{amsthm}
\usepackage{mathtools}
\usepackage{color}
\usepackage{amsfonts}
\newcommand{\bs}{\boldsymbol}

\newcommand{\sign}{{\rm sign}}
\newcommand{\two}{{\rm two}}

\newcommand{\mc}{\mathcal}
\newcommand{\cycle}{\Sigma_{\rm cyc}\ }
\newcommand{\interior}{{\rm int}\ }
\newcommand{\ext}{{\rm ext}}
\newcommand{\sumi}{\Sigma_i \ }
\newcommand{\projection}{\mathrm{Proj}^{-1}}
\newcommand{\set}{\mc{N}}
\newtheorem{theorem}{Theorem}
\newtheorem{lemma}{Lemma}

\newtheorem{corollary}{Corollary}
\usepackage{comment}
\usepackage{bm}

\makeatletter
\gdef\@copyrightpermission{
  \begin{minipage}{0.2\columnwidth}
   \href{https://creativecommons.org/licenses/by/4.0/}{\includegraphics[width=0.90\textwidth]{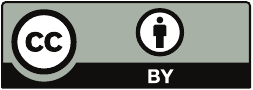}}
  \end{minipage}\hfill
  \begin{minipage}{0.8\columnwidth}
   \href{https://creativecommons.org/licenses/by/4.0/}{This work is licensed under a Creative Commons Attribution International 4.0 License.}
  \end{minipage}
  \vspace{5pt}
}
\makeatother

\setcopyright{ifaamas}
\acmConference[AAMAS '25]{Proc.\@ of the 24th International Conference
on Autonomous Agents and Multiagent Systems (AAMAS 2025)}{May 19 -- 23, 2025}
{Detroit, Michigan, USA}{Y.~Vorobeychik, S.~Das, A.~Nowé  (eds.)}
\copyrightyear{2025}
\acmYear{2025}
\acmDOI{}
\acmPrice{}
\acmISBN{}

\acmSubmissionID{7}

\title[Nash Equilibrium and Learning Dynamics in Three-Player Matching $m$-Action Games]{Nash Equilibrium and Learning Dynamics\\ in Three-Player Matching $m$-Action Games}

\subtitle{Extended Abstract}

\author{Yuma Fujimoto}
\affiliation{
  \institution{CyberAgent}\country{}
  \institution{University of Tokyo}\country{}
  \institution{Soken University}\country{}
  }
\email{fujimoto.yuma1991@gmail.com}

\author{Kaito Ariu}
\affiliation{
  \institution{CyberAgent}\country{}
  }
\email{kaito_ariu@cyberagent.co.jp}

\author{Kenshi Abe}
\affiliation{
  \institution{CyberAgent}\country{}
  }
\email{abekenshi1224@gmail.com}

\begin{abstract}
Learning in games discusses the processes where multiple players learn their optimal strategies through the repetition of game plays. The dynamics of learning between two players in zero-sum games, such as Matching Pennies, where their benefits are competitive, have already been well analyzed. However, it is still unexplored and challenging to analyze the dynamics of learning among three players. In this study, we formulate a minimalistic game where three players compete to match their actions with one another. Although interaction among three players diversifies and complicates the Nash equilibria, we fully analyze the equilibria. We also discuss the dynamics of learning based on some famous algorithms categorized into Follow the Regularized Leader. From both theoretical and experimental aspects, we characterize the dynamics by categorizing three-player interactions into three forces to synchronize their actions, switch their actions rotationally, and seek competition.
\end{abstract}


\newcommand{\BibTeX}{\rm B\kern-.05em{\sc i\kern-.025em b}\kern-.08em\TeX}

\begin{document}


\pagestyle{fancy}
\fancyhead{}


\maketitle 


\section{Introduction}
Learning in games considers that multiple agents independently learn their strategies to increase their utilities~\cite{fudenberg1998theory}. How to achieve the Nash equilibrium~\cite{nash1950equilibrium}, in which all agents are not motivated to change their strategies, is a key issue in learning in games. However, this issue is critical when their utility functions conflict with each other. The minimum example of such a conflict is matching pennies, where two agents have two actions, and their optimal actions are interdependent on each other, i.e., conflict.

To resolve this issue, the dynamics of agents' strategies have been enthusiastically studied in recent years~\cite{tuyls2005evolutionary, tuyls2006evolutionary, bloembergen2015evolutionary}. A representative method to analyze the dynamics is the learning algorithm named Follow the Regularized Leader (FTRL)~\cite{shalev2006convex, mertikopoulos2016learning, mertikopoulos2018cycles}, including several well-known algorithms, such as replicator dynamics~\cite{taylor1978evolutionary, friedman1991evolutionary, hofbauer1998evolutionary, borgers1997learning, hofbauer1998evolutionary, sato2002chaos} and gradient ascent~\cite{hofbauer1990adaptive, dieckmann1996dynamical, singh2000nash, zinkevich2003online}. The above matching pennies game attracts a lot of attention to discuss the dynamics, where their strategies draw a cycle around the Nash equilibrium~\cite{borgers1997learning, bloembergen2015evolutionary}. Such cycling dynamics are also observed in some variations of games~\cite{hofbauer1998evolutionary, singh2000nash, bailey2019multi} derived from the matching pennies and are understood by a conserved quantity~\cite{piliouras2014persistent, mertikopoulos2018cycles, bailey2019multi}; The existence of the conserved quantity is universally seen in the FTRL. As a topic developing recently, various algorithms are proposed and achieve the convergence to the Nash equilibrium (called last-iterate convergence), especially in games that usually result in cycles~\cite{bowling2000convergence, anagnostides2022last, fujimoto2024memory}. This convergence is discussed based on Lyapunov functions, corresponding to the distance from the Nash equilibrium. The Lyapunov functions are proven to decrease with time. To summarize, analyzing the cycling dynamics of agents' strategies matters in learning in games, where the matching pennies game is a keystone.

Despite such a thorough understanding of two-player games, three-player games are difficult to understand in general. Classically, this difficulty is seen in Jordan's game~\cite{jordan1993three}, which is a three-player version of matching pennies. In this game, the divergence from the equilibrium is observed~\cite{gaunersdorfer1995fictitious, mccabe2000experimental, shamma2005dynamic, mealing2015convergence}; the distance from the equilibrium is not a conserved quantity. This divergence is not seen in matching pennies. Like this, the motivation to study complex dynamics in learning in three-player games is established~\cite{kleinberg2011beyond, nagarajan2018three}. However, these studies consider poly-matrix games that can be divided into two-player interactions. Thus, how three-player interaction, which cannot be represented as a polymatrix game, changes the learning dynamics is still unclear. For games other than matching pennies, such a three-player interaction complicates the learning dynamics in rock-paper-scissors~\cite{sato2005stability} and social dilemma~\cite{akiyama2000dynamical}. In addition, the Nash equilibria of three-player games are hard to fully analyze~\cite{daskalakis2005three}, even when the games are zero-sum. In other words, the Nash equilibria are expected to be more complex in three-player games than in two-player games. Indeed, it is difficult to analyze the Nash equilibria in Kuhn poker~\cite{szafron2013parameterized} and repeated prisoner's dilemma~\cite{murase2018seven} and the correlated equilibria in rock-paper-scissors~\cite{grant2023correlated}. It should also be noted that such an equilibrium analysis is deeply tied to the development of AI in table games. Game AIs have developed faster in two-player table games, such as Go~\cite{silver2017mastering}, while they are still developing in more than two-player table games, including Poker~\cite{brown2019superhuman}, Mahjong~\cite{li2020suphx}, and Diplomacy~\cite{paquette2019no}. To summarize, understanding how three-player interactions change the Nash equilibria and learning dynamics is important to gain insight into the three-player game.

This study extends the ordinary matching pennies game to a three-player general $m$-action version and names it Three-Player Matching $m$-Action ($m$-3MA) game. Depending on its payoff matrix, this $m$-3MA becomes a general-sum game and thus has a complex Nash equilibrium structure. Despite such complexity, we fully analyze the Nash equilibria in $m$-3MA games. Furthermore, we introduce the continuous-time FTRL algorithm. Next, we analyze the dynamics of this algorithm in $m$-3MA for several representative regularizers, i.e., the entropic and Euclidean ones. We observe that the dynamics provide various behaviors depending on the parameters of three-player interaction: cycle around the Nash equilibria, convergence to it, and divergence from it with a heteroclinic cycle. To understand these diverse dynamics, we introduce a novel Lyapunov function, which is the degree of synchronization among three players' actions, and prove that this quantity captures the global behavior of the dynamics. We also interpret the global behavior depending on the parameters of three-player interactions.

\section{Preliminary}
\subsection{Three-Player Matching $m$-Action Games}
We now assume a situation where one's action is advantageous under a certain pair of the others' actions but is disadvantageous under another pair. Thus, the property of the matching pennies game is inherited. Based on this situation, we now formulate Three-Player Matching $m$-Action ($m$-3MA) games.

\begin{figure}[h!]
    \centering
    \includegraphics[width=1.0\hsize]{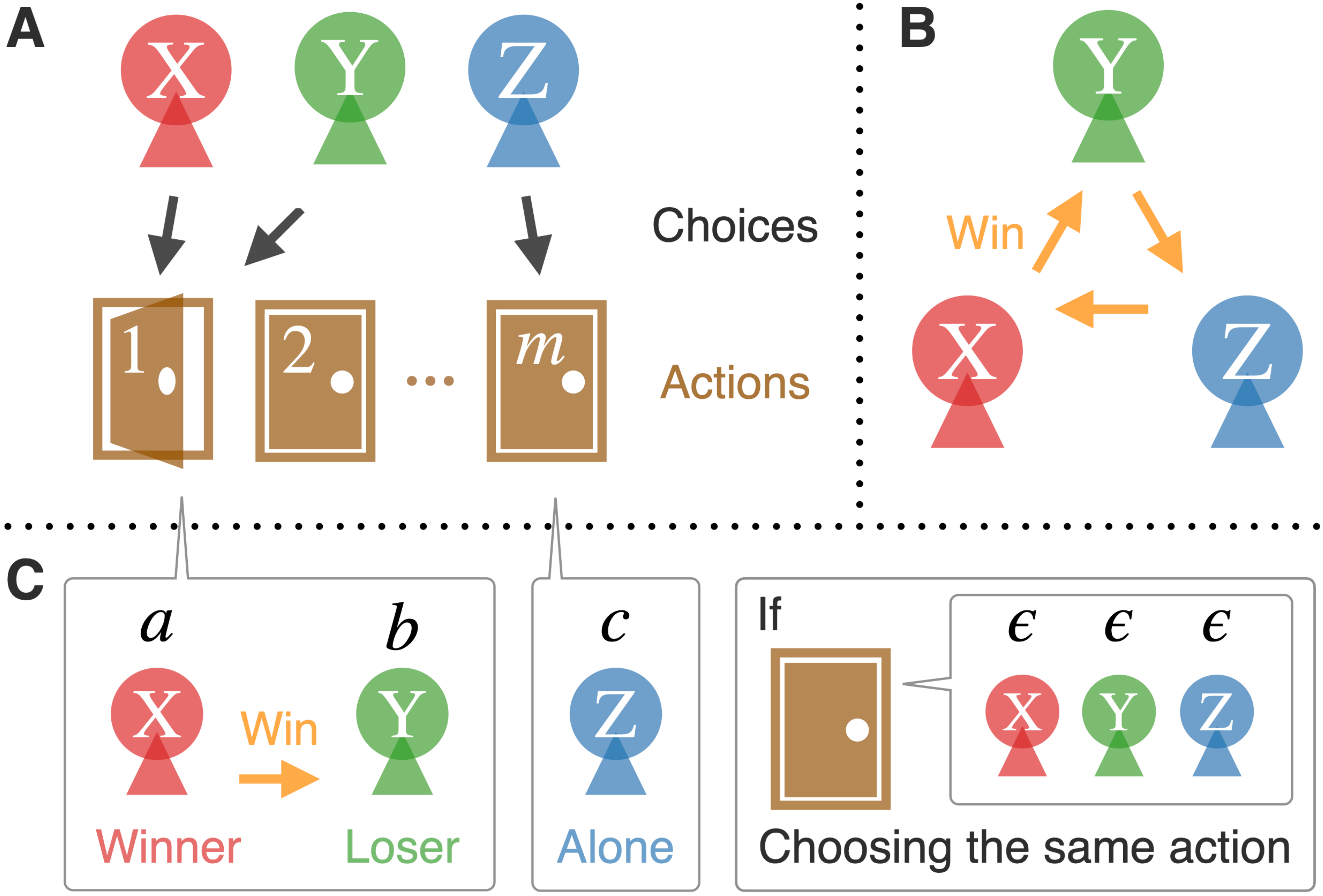}
    \caption{{\bf A}. Three players, X, Y, and Z, independently choose their actions. Players who choose the same action play the game together. {\bf B}. In the game, players have a three-way deadlock relationship. X, Y, and Z are advantageous to Y, Z, and X, respectively. {\bf C}. The three players receive their own scores as a result of their action choices. When two of the three players (X and Y in the left panel) choose the same action, the winner's score is $a$, while the loser's score is $b$, following the three-way deadlock relationship. An isolated player (Z in the center panel) who chooses a different action from others receives a score of $c$. If all three players choose the same action (in the right panel), they receive scores of $\epsilon$. Here, we assume $b<c<a$ and $b<\epsilon<a$.}
    \label{F01}
\end{figure}

Let X, Y, and Z denote three players. Every round, they independently determine their actions from the same $m$-action set, $\mc{A}=\{a_1,\cdots,a_m\}$ (see Fig.~\ref{F01}-A). Players who choose the same action interact with each other. This interaction follows a three-way deadlock relationship among them (see Fig.~\ref{F01}-B): X wins Y, Y wins Z, but Z wins X. They receive their scores (see Fig.~\ref{F01}-C). When only two of them interact, the winner and loser are determined following the three-way deadlock relationship, and the winner's and loser's scores are $a$ and $b$ (see the left panel of Fig.~\ref{F01}-C), respectively. Players who chose a different action from the others receive the default payoff of $c$ (see the center). If all three players take the same action, they commonly receive the scores of $\epsilon$ (see the right). Here, we assume that the winner's and loser's scores are highest and lowest, respectively, i.e., $b<c<a$ and $b<\epsilon<a$.

We also formulate their mixed strategies and payoffs. Let $0\le x_i\le 1$ denote the probability that X chooses action $a_i$. X's strategy is defined as $\bs{x}:=(x_1,\cdots,x_m)\in\Delta^{m-1}$ (the $m-1$ dimensional simplex). Similarly, Y's and Z's strategies are denoted by $\bs{y}\in\Delta^{m-1}$ and $\bs{z}\in\Delta^{m-1}$, respectively. When players follow such strategies, X's expected payoff is given by
\begin{align}
    u(\bs{x},\bs{y},\bs{z})&=\epsilon\sum_{i}x_iy_iz_i+a\sum_{i}x_iy_i\bar{z}_i
    \nonumber\\
    &+b\sum_{i}x_i\bar{y}_iz_i+c\sum_{i}x_i\bar{y}_i\bar{z}_i,
\end{align}
where we defined $\bar{\mc{X}}:=1-\mc{X}$ for arbitrary variable $\mc{X}$. Y's and Z's expected payoffs are also described as $u(\bs{y},\bs{z},\bs{x})$ and $u(\bs{z},\bs{x},\bs{y})$, respectively.

In learning in games, the gradient of this payoff is often important, calculated as
\begin{align}
    &\frac{\partial u(\bs{x},\bs{y},\bs{z})}{\partial x_i}
    \nonumber \\
    &=\epsilon y_iz_i+ay_i\bar{z}_i+b\bar{y}_iz_i+c\bar{y}_i\bar{z}_i,
    \nonumber \\
    &=(\epsilon-a-b+c)y_iz_i+(a-c)y_i+(b-c)z_i+c,
    \nonumber \\
    &=(\alpha-\gamma)y_iz_i+\frac{\beta+\gamma}{2}y_i+\frac{-\beta+\gamma}{2}z_i+c,
    \nonumber \\
    &=:f(y_i,z_i),
\end{align}
where we define $\alpha:=\epsilon-c$, $\beta:=a-b>0$, and $\gamma:=a+b-2c$. In this payoff gradient, $c$ is the offset and thus negligible. Thus, we can characterize the $m$-3MA games by the three parameters of $\alpha$, $\beta$, and $\gamma$.

\subsection{Real-world application}
This $m$-3MA game is applicable to some real-world situations. For example, imagine that three companies (denoted as X, Y, and Z, respectively) are selling their products in $m$ markets. Next, consider the case where X's product is purchased more than Y's by buyers, Y's than Z's, but Z's than X's. Although this case may seem contradictory, it can occur depending on individual preferences and holds significant relevance in the context of social choice~\cite{arrow2012social, tversky1969intransitivity}. In this example, in order to increase its benefit, each company would control the supply of its products to the market. This control process has often been modeled as evolutionary~\cite{taylor1978evolutionary} and adaptive dynamics~\cite{hofbauer1990adaptive}. These dynamics correspond to the replicator dynamics and gradient ascent, both of which are included in the following continualized FTRL algorithms.

Similar games to $m$-3MA are often studied even in other fields. For example, in ecological systems, the three-way deadlock relationship between three species (X, Y, and Z) is often observed, where X eliminates Y, Y does Z, and Z to X (e.g., the relationship among X: Japanese honeybee, Y: Japanese hornet, and Z: European honeybee~\cite{tainaka2000physics}). These three species would compete for their territory in $m$ patches (corresponding to actions in our study). Their population dynamics are described as the replicator dynamics in standard. In physics, a huge amount of papers have numerically simulated the complex dynamics of their populations (e.g.,~\cite{reichenbach2007mobility}). To summarize, the motivation to study three-player games like ours is shared in a wide range of fields.

\section{Nash Equilibrium}
The Nash equilibrium of $m$-3MA is defined as the set of strategies $(\bs{x}^*,\bs{y}^*,\bs{z}^*)$ that satisfy
\begin{align}
    \begin{cases}
        \bs{x}^*\in\arg\max_{\bs{x}}u(\bs{x},\bs{y}^*,\bs{z}^*) \\
        \bs{y}^*\in\arg\max_{\bs{y}}u(\bs{y},\bs{z}^*,\bs{x}^*) \\
        \bs{z}^*\in\arg\max_{\bs{z}}u(\bs{z},\bs{x}^*,\bs{y}^*) \\
    \end{cases}.
\end{align}

It is difficult to derive equilibrium in three-player games~\cite{daskalakis2005three}, and indeed, there are few successful studies~\cite{szafron2013parameterized, grant2023correlated}. Nevertheless, we can fully analyze all the Nash equilibria in Thm.~\ref{thm_Nash} in Sec.~\ref{sec_Full}. Because the theorem is based on rigorous but complicated calculations, its visualization and brief interpretation are provided in Sec.~\ref{sec_Visualization}.

\subsection{Full Analysis} \label{sec_Full}
We first introduce three types of strategies consisting of the Nash equilibrium. The first is the uniform-choice equilibrium $\set_{\rm U}(m):=\{\bs{1}/m\}$, where $\bs{1}$ is defined as the $m$-dimensional all-ones vector. In this equilibrium, the player chooses all the actions at perfectly random. The second is the pure-strategy equilibria $\set_{\rm P}(m):=\{\bs{e}_1,\cdots,\bs{e}_m\}$, where $\bs{e}_i$ is the unit vector for the $i$-th axis in the $m$-dimensional space. These equilibria mean that the player deterministically chooses an action (uses a pure strategy). The third is the double roots equilibria;
\begin{align}
    \set_{\rm DR}(m):=\begin{cases}
        \{\sigma(\bs{x}_{{\rm DR};k})\;|\;\sigma\in S_m,k\le 1/(2x_{\ext})\} \\
        \hspace{3cm} (x_{\ext}>1/m) \\
        \{\sigma(\bs{x}_{{\rm DR};k})\;|\;\sigma\in S_m,k\ge 1/(2x_{\ext})\} \\
        \hspace{3cm} (x_{\ext}<1/m) \\
    \end{cases},
\end{align}
where we used the permutation function $\sigma$ for $m$-dimensional symmetry group $S_m$, and also used
\begin{align}
    x_{\ext}&:=\frac{\gamma}{2(\gamma-\alpha)}, \\
    \delta&:= \frac{mx_{\ext}-1}{m-2k}, \\
    x_{+}&:=x_{\ext}+\delta, \\
    x_{-}&:=x_{\ext}-\delta, \\
    \bs{x}_{{\rm DR};k}&:=(\underbrace{x_{+},\cdots,x_{+}}_{k},\underbrace{x_{-},\cdots,x_{-}}_{m-k}).
\end{align}
These equilibria are very complex because the player chooses each action with two different probabilities $x_{+}$ and $x_{-}$.

In addition, for the range of $A=\{2,\cdots,m-1\}$, we define
\begin{align}
    \set'_{\rm U}(m)&:= \set_{\rm U}(m)\cup\left(\bigcup_{m'\in A}\projection(\set_{\rm U}(m'))\right),
\end{align}
where $\projection$ denotes an inverse projection map from $m'(<m)$- to $m$-dimensional vector set with arbitrary permutations. Thus, $\set'_{\rm U}(m)$ contains all the uniform-choice equilibria of $m'(<m)$-action games. Here, we split $A$ into $A^{(-)}:=\{2,\cdots,\lfloor1/(2x_{\ext})\rfloor\}$ and $A^{(+)}:=\{\lceil 1/(2x_{\ext})\rceil,\cdots,m-1\}$ as $A=A^{(-)}\cup A^{(+)}$. Then, $\set'^{(-)}_{\rm U}(m)$ (resp. $\set'^{(+)}_{\rm U}(m)$) denotes that the union set in the range of $m'\in A^{(-)}$ (resp. $A^{(+)}$). $\set'_{\rm DR}(m)$ is similarly defined.

Using the above definition, we derive the Nash equilibria as follows.

\begin{figure*}[h!]
    \centering
    \includegraphics[width=1.0\hsize]{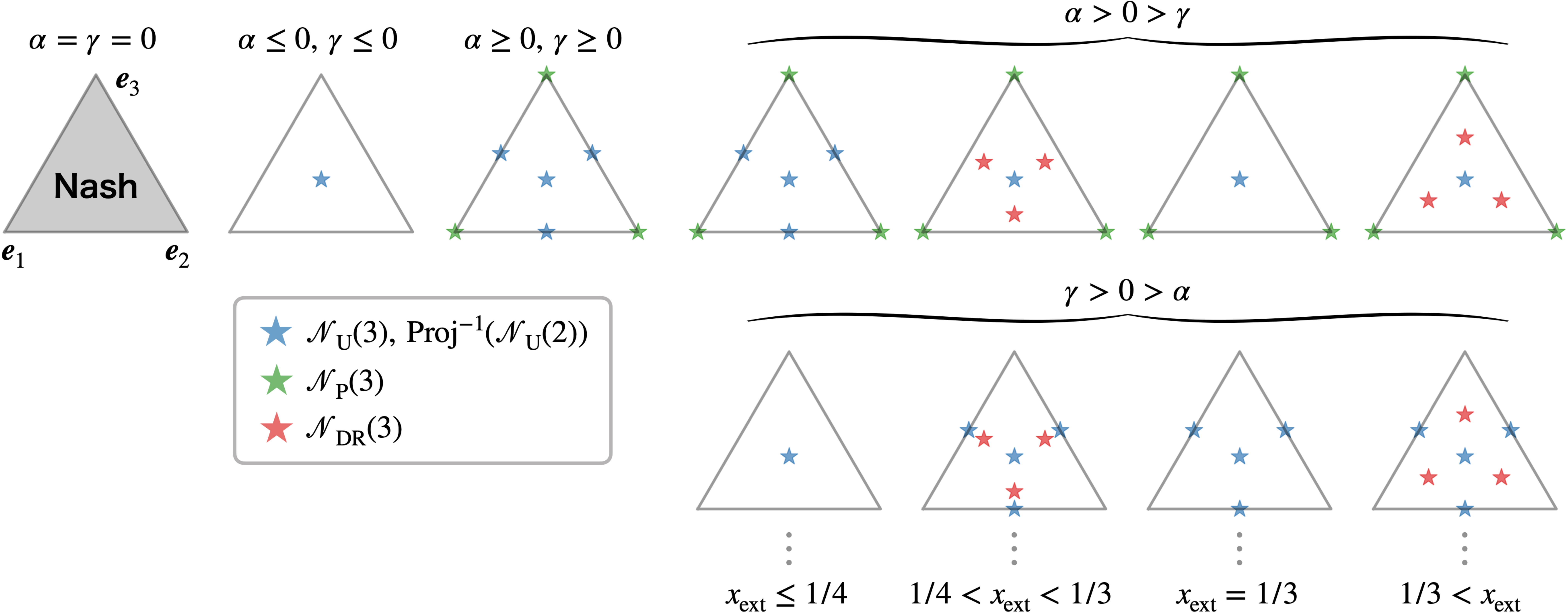}
    \caption{The Nash equilibrium in $m$-3MA with $m=3$. This Nash equilibrium crucially changes depending on $\alpha$ and $\gamma$. Each panel shows the simplex $\Delta^{2}$ of one's strategy. The edges of triangles indicate the pure strategies, i.e., $\bs{x}=\bs{y}=\bs{z}=\bs{e}_i$, where all the players choose only action $i$. The blue stars represent $\set_{\rm U}(3)$ (and $\projection(\set_{\rm U}(2))$). The green stars represent $\set_{\rm P}(3)$. Finally, the red stars represent $\set_{\rm DR}(3)$. Here, note that since we consider the case of $m=3$, $\projection(\set_{\rm DR}(2))=\emptyset$ always hold. Here, the positions of $\set_{\rm DR}$ change depending on $x_{\ext}$, which is determined by $\alpha$ and $\gamma$.}
    \label{F02}
\end{figure*}

\begin{theorem}[Nash equilibrium solution]
\label{thm_Nash}
For any Nash equilibrium $(\bs{x}^*,\bs{y}^*,\bs{z}^*)$, $\bs{x}^*=\bs{y}^*=\bs{z}^*$ holds, and the set of one's strategies is given by
\begin{align}
    \begin{cases}
        \alpha=\gamma=0 &\Rightarrow \Delta^{m-1} \\
        0\ge\alpha, 0\ge\gamma &\Rightarrow \set_{\rm U}(m) \\
        \alpha>0>\gamma &\Rightarrow \set_{\rm P}(m)\cup\set'^{(-)}_{\rm U}(m)\cup\set_{\rm DR}(m) \\
        \alpha\ge 0, \gamma\ge 0 &\Rightarrow \set_{\rm P}(m)\cup\set'_{\rm U}(m) \\
        \gamma>0>\alpha &\Rightarrow \set_{\rm U}'^{(+)}(m)\cup\set'_{\rm DR}(m) 
    \end{cases}.
\end{align}
\end{theorem}

See Appendix.~\ref{proof_thm_Nash}-\ref{proof_lem_Boundary} for its complete proof.

\textsc{Proof Sketch}. Step 1 (Lemma~\ref{lem_Symmetry}): We prove $\bs{x}^*=\bs{y}^*=\bs{z}^*$ by contradiction. In other words, we derive a contradiction from the assumption of $\bs{x}^*\neq\bs{y}^*\neq\bs{z}^*\neq\bs{x}^*$ or $\bs{x}^*=\bs{y}^*\neq\bs{z}^*$. This contradiction is proved because there is no ordering among $f(y_i^*,z_i^*)$, $f(z_i^*,x_i^*)$, and $f(x_i^*,y_i^*)$. Step~2 (Lemma~\ref{lem_Interior} and~\ref{lem_Boundary}): Under the condition of $\bs{x}^*=\bs{y}^*=\bs{z}^*$, we find the set of one's equilibrium strategies. Both cases when a strategy exists in the interior or on the boundary of the strategy spaces are considered. \qed

\subsection{Visualization and Interpretation} \label{sec_Visualization}
To visualize Thm.~\ref{thm_Nash}, Fig.~\ref{F02} illustrates all the cases of the Nash equilibria. As this figure shows, $m$-3MA provides diverse and complicated Nash equilibrium structures depending on $\alpha$ and $\gamma$. The Nash equilibria are especially complicated when $\alpha$ and $\gamma$ conflict (i.e., $\alpha>0>\gamma$ or $\gamma>0>\alpha$). For further interpretation of the Nash equilibria, the following corollary summarizes the main properties of the equilibria.


\begin{corollary}[Main properties of the Nash equilibria]
\label{Cor_Main}
First, the following property always holds.
\begin{itemize}
\item {\bf (Player symmetry)} For any Nash equilibrium, all three players take the same strategy, i.e., $\bs{x}^*=\bs{y}^*=\bs{z}^*$.
\end{itemize}
Furthermore, the region of the Nash equilibria has the following properties.
\begin{itemize}
\item {\bf (Neutral equilibria)} When $\alpha=\gamma=0$, all the strategies in the simplex $\Delta^{m-1}$ can be the Nash equilibria.
\item {\bf (Pure-strategy equilibria)} $\set_{\rm P}(m)=\{\bs{e}_1,\cdots,\bs{e}_m\}$ are the Nash equilibrium strategies, if and only if $\alpha\ge 0$.
\item {\bf (Uniform-choice equilibrium)} $\set_{\rm U}(m)=\{\bs{1}/m\}$ is always the Nash equilibrium strategy.
\end{itemize}
\end{corollary}

\begin{proof}
These properties are immediately derived by Thm.~\ref{thm_Nash}.
\end{proof}

We now interpret these properties of the Nash equilibria. {\bf (Neutral equilibrium)} First, the game parameters of $\alpha$ and $\gamma$ represent interactions among three players (see Sec.~\ref{sec_Experimental} and Sec.~\ref{sec_Three} for details of $\alpha$ and $\gamma$, respectively). The games with $\alpha=\gamma=0$ are classified as a zero-sum game that satisfies monotonicity. Thus, no three-player interaction works, and a continuous region can be equilibrium. {\bf (Pure-strategy equilibria)} The games with $\alpha>0$ have a positive three-player interaction, where positive payoffs are generated when three players interact. In such games, the players are motivated to synchronize their actions. Thus, the states in which the three players perfectly synchronize their action choices, i.e., the pure-strategy equilibria, can be an equilibrium. {\bf (Uniform-choice equilibrium)} When the other two players take all actions at perfectly random, i.e., use $\bs{1}/m$, all the actions are equivalent for a player. Thus, the uniform-choice equilibrium $\bs{x}^*=\bs{y}^*=\bs{z}^*=\set_{\rm U}(m)$ is always the Nash equilibrium independent of $\alpha$.

\section{Learning algorithm}
This section considers the continuous-time Follow the Regularized Leader (FTRL), which is formulated as
\begin{align}
    \bs{x}&=\bs{q}(\bs{x}^{\dagger}),
    \label{FTRL1}\\
    \dot{\bs{x}}^{\dagger}&=\frac{\partial u}{\partial \bs{x}},
    \label{FTRL2}\\
    \bs{q}(\bs{x}^{\dagger})&=\arg\max_{\bs{x}}\left\{\bs{x}^{\dagger}\cdot\bs{x}-h(\bs{x})\right\}.
    \label{FTRL3}
\end{align}
Here, $h(\bs{x})$ is ``regularizer'', a penalty term in projecting the updated strategy back to its strategy space. Several representative examples are the entropic regularizer $h(\bs{x})=\bs{x}\cdot\log\bs{x}$ and the Euclidean regularizer $h(\bs{x})=\|\bs{x}\|^2/2$. The following lemma shows that these regularizers provide the replicator dynamics and gradient ascent, based on the result in~\cite{mertikopoulos2018cycles}.

\begin{lemma}[Replicator dynamics and gradient ascent]
\label{lem_RDGA}
{\bf (Replicator dynamics)} In $m$-3MA, the continuous-time FTRL with $h(\bs{x})=\bs{x}\cdot\log\bs{x}$ are calculated as
\begin{align}
    \dot{x}_i=&x_i\left(f(y_i,z_i)-\sum_ix_if(y_i,z_i)\right).
    \label{ReplicatorDynamics}
\end{align}

{\bf (Gradient ascent)} In the interior of the players' strategy spaces, the continuous-time FTRL with $h(\bs{x})=\|\bs{x}\|^2/2$ are calculated as
\begin{align}
    \dot{x}_i=&f(y_i,z_i)-\frac{1}{m}\sum_i f(y_i,z_i).
    \label{GradientAscent_interior}
\end{align}
\end{lemma}

\section{Learning Dynamics}
Let us consider learning dynamics by continuous-time FTRL. In Sec.~\ref{sec_Characterization}, two quantities, $G$ and $V$, are introduced to characterize the learning dynamics. In Sec.~\ref{sec_Two}, we theoretically analyze the global behavior of the learning dynamics by using $G$ and $V$. Finally, in Secs.~\ref{sec_Two} and \ref{sec_Three}, we simulate the learning dynamics for $m=2$ and $m>2$ and experimentally demonstrate how game parameters, i.e., $\alpha$, $\beta$, and $\gamma$, contribute to the dynamics.

\subsection{Characterization of $m$-3MA} \label{sec_Characterization}
To investigate learning dynamics given by the continuous-time FTRL, we introduce $G$ and $V$ as
\begin{align}
    G(\bs{x}^{\dagger},\bs{y}^{\dagger},\bs{z}^{\dagger})&:=\cycle \max_{\bs{x}}\{\bs{x}^{\dagger}\cdot\bs{x}-h(\bs{x})\} - \bs{x}^{\dagger}\cdot\bs{1}/m,\\
    V(\bs{x},\bs{y},\bs{z})&:=\sumi x_iy_iz_i-1/m^2.
\end{align}
Here, $\cycle$ indicates the cyclic sum for three players. In other words, $\cycle \mc{F}(\bs{x})=\mc{F}(\bs{x})+\mc{F}(\bs{y})+\mc{F}(\bs{z})$ holds for arbitrary function $\mc{F}$. We now explain the meanings of $G$ and $V$. First, $G$ is known to be conserved under zero-sum games~\cite{mertikopoulos2018cycles} and corresponds to Kullback-Leibler divergence from the uniform-choice equilibrium. Next, $V$ means the probability that all three players choose the same action, in other words, the degree of synchronization of their action choices. The following lemma gives the basics of this $V$;

\begin{lemma}[Properties of $V$]
\label{lem_Properties}
$V$ takes its maximum value if and only if $\bs{x}=\bs{y}=\bs{z}\in\set_{\rm P}(m)$ and its minimum value if and only if $x_i=0$, $y_i=0$, or $z_i=0$ hold for each $i\in\{1,\cdots,m\}$.
\end{lemma}

\begin{proof}
For the maximum value condition, $\sumi x_iy_iz_i\le \sum_i x_iy_i\le 1$ trivially holds. The second equation holds if and only if $\bs{x}=\bs{y}\in\set_{\rm P}(m)$ by the definition of an inner product. The first equation further holds if and only if $\bs{x}=\bs{y}=\bs{z}\in\set_{\rm P}(m)$. For the minimum value condition, $0\le \sumi x_iy_iz_i$ trivially holds. Here, this equation holds if and only if $x_iy_iz_i=0\Leftrightarrow (x_i=0)\lor(y_i=0)\lor(z_i=0)$ for all $i\in\{1,\cdots,m\}$. \end{proof}

\subsection{Analysis and Experiments of Two-Action Games} \label{sec_Two}
\subsubsection{Theoretical Analysis}
We now consider $m$-3MA with the case of $m=2$. Since $m=2$ holds, $x_2=1-x_1$, $y_2=1-y_1$, and $z_2=1-z_1$ hold so that we can describe the learning dynamics by the three variables of $(x_1,y_1,z_1)$. The dynamics given by the FTRL algorithm in the game are independent of $\gamma$, as proved in the following.

\begin{lemma}[Simplified dynamics for $m=2$]
\label{lem_Dynamics_two}
When $m=2$, the dynamics in Lem.~\ref{lem_RDGA} are calculated as
\begin{align}
    \dot{x}_1&=w(x_1)\{\alpha(y_1+z_1-1)+\beta(y_1-z_1)\},
    \label{FTRL_two-action} \\
    w(x_1)&=\begin{cases}
        x_1(1-x_1) & (h(\bs{x})=\bs{x}\cdot\log\bs{x}) \\
        1/2 & (h(\bs{x})=\|\bs{x}\|^2/2) \\
    \end{cases},
\end{align}
independent of $\gamma$.
\end{lemma}

See Appendix.~\ref{proof_lem_Dynamics} for its complete proof.

\textsc{Proof Sketch}. Since $x_2=\bar{x}_1$, $y_2=\bar{y}_1$, and $z_2=\bar{z}_1$ hold in two-action games, the dynamics in Lem.~\ref{lem_RDGA} are rewritten in a simpler form by direct calculation. \qed

Under the entropic and Euclidean regularizers, the time changes of these $V$ and $G$ are characterized by the following two theorems.

\begin{theorem}[Monotonicity of $V$]
\label{thm_Monotonicity}
In $m$-3MA with $m=2$, the FTRL algorithm with the entropic regularizer $h(\bs{x})=\bs{x}\cdot\log\bs{x}$ and Euclidean reguralizer $h(\bs{x})=\|\bs{x}\|^2/2$ gives
\begin{align}
    \sign(\dot{V})=\sign(\alpha),
\end{align}
in the interior of strategy space other than $\bs{x}=\bs{y}=\bs{z}=\bs{1}/m$.
\end{theorem}

See Appendix.~\ref{proof_thm_Monotonicity} for its complete proof.

\textsc{Proof Sketch}. We calculate the dynamics of $V$ by using Eq.~\eqref{FTRL_two-action}. The two terms of $\alpha$ and $\beta$ contribute to the dynamics. Here, the term of $\beta$ disappears by the cyclic symmetry among the three players. The term of $\alpha$ is always positive, meaning that the signs of $\dot{V}$ and $\alpha$ are the same. \qed

\begin{theorem}[Connection between $G$ and $V$]
\label{thm_Connection}
In $m$-3MA with $m=2$, the FTRL algorithm with the entropic regularizer $h(\bs{x})=\bs{x}\cdot\log\bs{x}$ and Euclidean reguralizer $h(\bs{x})=\|\bs{x}\|^2/2$ gives $\dot{G}= 2\alpha V$ in the interior of strategy space other than $\bs{x}=\bs{y}=\bs{z}=\bs{1}/m$.
\end{theorem}

See Appendix.~\ref{proof_thm_Connection} for its complete proof.

\begin{figure*}[h!]
    \centering
    \includegraphics[width=0.85\hsize]{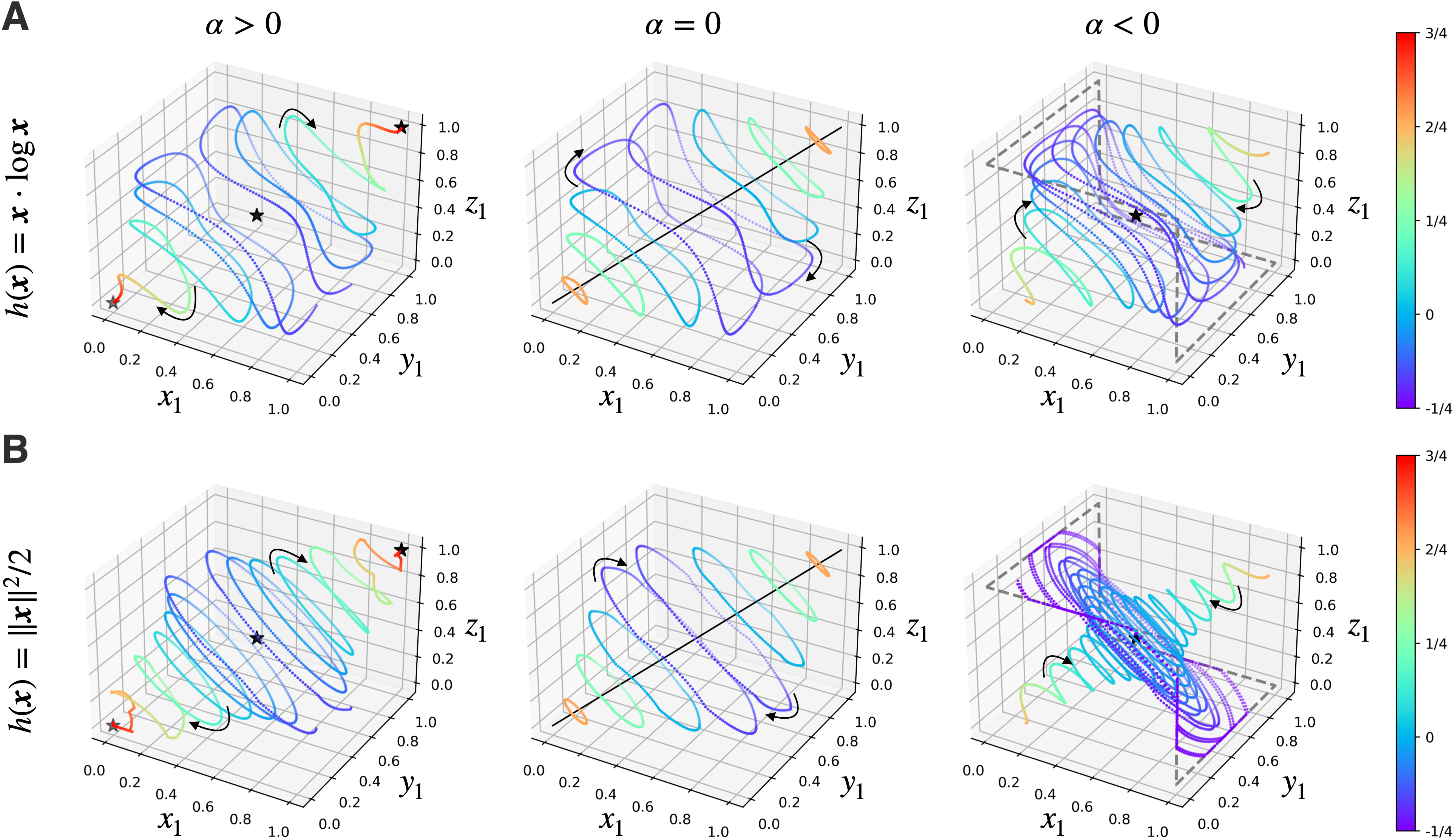}
    \caption{{\bf A}. The dynamics of FTRL with the entropic regularizer in $m$-3MA with $m=2$. The dynamics are output by the fourth-order Runge-Kutta method with the step-size of $2\times 10^{-2}$ in all the panels. We also commonly set $(a,b,c)=(1,-1,0)$, in other words, $\beta=2$. In the left, center, and right panels, we set $\epsilon(=\alpha)=0.1$, $0$, $-0.1$, respectively. The red, green, and blue lines indicate the time series of $x_i$, $y_i$, and $z_i$, respectively, while the solid, broken, and dotted lines indicate $i=1$, $2$, and $3$, respectively. The solid black line indicates the time series of $V$. The initial strategies in each panel are randomly sampled from the strategy simplexes. In the left and right panels, the Nash equilibria are plotted by the black star. In the center panel, all the points on the black solid line are the Nash equilibria. The broken line in the right panel is the set of states satisfying the minimum $V$ condition. {\bf B}. The dynamics of FTRL with the Euclidean regularizer. The method and simulation parameters are the same as panel A.}
    \label{F03}
\end{figure*}

\textsc{Proof Sketch}. The dynamics of $G$ are calculated by using Eq.~\eqref{FTRL_two-action} and depend on both $\alpha$ and $\beta$. Here, the term of $\beta$ disappears again by the cyclic symmetry among the three players. The term of $\alpha$ corresponds to $2V$. \qed

These two theorems explain the global behavior of the learning dynamics as follows.

\begin{corollary}[Global behavior of dynamics]
\label{cor_Global}
In $m$-3MA with $m=2$, the continuous-time FTRL gives the following properties except for when the trajectory converges to the uniform-choice equilibrium.
\begin{itemize}
\item When $\alpha=0$, both $G$ and $V$ are conserved in the trajectory.
\item When $\alpha>0$, the trajectory asymptotically converges to the states of maximum $V$, i.e., either of the fixed points.
\item When $\alpha<0$, the trajectory asymptotically converges to the states of minimum $V$, i.e., the heteroclinic cycle.
\end{itemize}
\end{corollary}

\begin{proof}
This is immediately proved by Lem.~\ref{lem_Properties}, Thm.~\ref{thm_Monotonicity}, and Thm.~\ref{thm_Connection}.
\end{proof}

\subsubsection{Experimental Understanding} \label{sec_Experimental}
We numerically demonstrate the learning trajectories in $m$-3MA with $m=2$. Fig.~\ref{F03}-A and B show the learning dynamics by continuous-time FTRL with the entropic (replicator dynamics) and Euclidean (gradient ascent) regularizers, respectively. Each figure plots the dynamics in the three cases of $\alpha>0$ (left), $\alpha=0$ (center), and $\alpha<0$ (right). One can see some differences between Fig.~\ref{F03}-A and B. The trajectory of the replicator dynamics has a distorted shape and always stays in the interior of the strategy space. On the other hand, the trajectory of the gradient ascent has a circular shape and often stays on the boundary of the strategy space. Regardless of such a difference, the trajectory commonly shows the properties given by Cor.~\ref{cor_Global} as follows.


\paragraph{Cycling behavior ($\alpha=0$):} First, we see the case of $\alpha=0$. Thm.~\ref{Cor_Main} shows that the Nash equilibria are given by the diagonal line of $\bs{x}^*=\bs{y}^*=\bs{z}^*\in\Delta^{1}$. Then, the learning dynamics always give cycling behavior around this diagonal line. As Cor.~\ref{cor_Global} proves, both $G$ and $V$ are invariant. Indeed, we can see that $V$ is constant in each trajectory (plotted by the same color), while $G$, the distance from the uniform-choice equilibrium in the center point, also seems constant in each trajectory.

\paragraph{Convergence to the pure-strategy equilibria ($\alpha>0$):} In $\alpha>0$, the Nash equilibria are given by $\bs{x}^*=\bs{y}^*=\bs{z}^*\in\set_{\rm P}(2)\cup\set_{\rm U}(2)=\{\bs{e}_1,\bs{e}_2,\bs{1}/2\}$. As Cor.~\ref{cor_Global} proves, the learning dynamics converge to either one of the pure-strategy equilibria $\bs{e}_1$ or $\bs{e}_2$ depending on their initial condition. Indeed, we can see that $V$ monotonically increases in each trajectory
(the plotted color changes into red monotonically).

\begin{figure*}[h!]
    \centering
    \includegraphics[width=0.8\hsize]{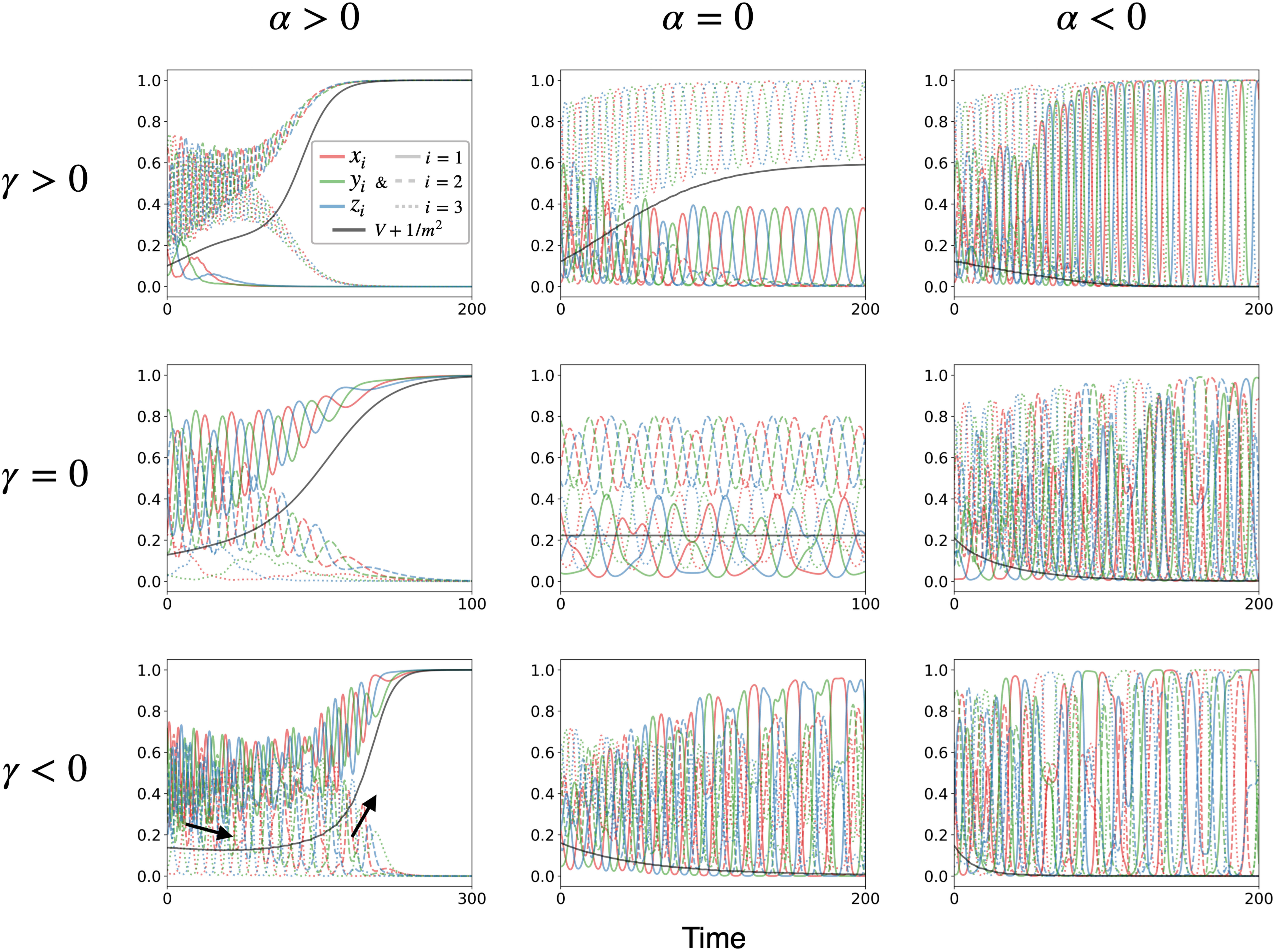}
    \caption{The dynamics of FTRL with the entropic regularizer. The dynamics are output by the fourth-order Runge-Kutta method with the step-size of $2\times 10^{-2}$ in all the panels. For $\alpha>0$, $=0$, and $<0$, we set $(\epsilon,c)=(0.1,0)$, $(0,0)$, and $(-0.1,0)$, respectively. For $\gamma>0$, $=0$, and $<0$, we set $(a,b,c)=(1.1,-0.9,0)$, $(1,-1,0)$, and $(0.9,-1.1,0)$, respectively.}
    \label{F04}
\end{figure*}

\paragraph{Heteroclinic cycles ($\alpha<0$):} In $\alpha<0$, the Nash equilibrium is only at $\bs{x}^*=\bs{y}^*=\bs{z}^*\in\set_{\rm U}(2)=\{\bs{1}/2\}$. Interestingly, the learning dynamics do not reach this Nash equilibrium but converge to the heteroclinic cycle of $(x_1,y_1,z_1)=(0,0,1)\to(0,1,1)\to(0,1,0)\to(1,1,0)\to(1,0,0)\to(1,0,1)\to(0,0,1)\to\cdots$. This means that as each player cyclically changes his/her action, he/she more biasedly chooses the action. In this heteroclinic cycle, $V$ takes its minimum value, and thus it is observed that Cor.~\ref{cor_Global} holds. The uniform-choice equilibrium $\bs{1}/2$ is unstable and cannot be reached from almost all the initial conditions.

\paragraph{$\alpha$ is the force of synchronization:} Let us explain the convergence in $\alpha>0$ and the divergence to the heteroclinic cycle in $\alpha<0$. Consider whether a player should choose the same action as the two others. If choosing, the player obtains $\epsilon$. Otherwise, $c$. In $\alpha>0\Leftrightarrow \epsilon>c$, the three players prefer to synchronize their actions more, eventually concentrating on choosing either action. On the other hand, in $\alpha<0\Leftrightarrow \epsilon<c$, the players prefer to desynchronize their actions.

\paragraph{$\beta$ is the force of rotation:} We also interpret $\beta=a-b$. Here, $a(>0)$ is the score for a winner. Thus, the larger $a$ is, the more quickly players are motivated to learn an advantageous action. On the other hand, $b(<0)$ is the score for a loser. The smaller $b$ means that players quickly learn to escape from being exploited. Because three players have a three-way deadlock relationship, their strategies show rotation. $\beta$ indicates the force of rotation.

\subsection{Experiments of More than Two-Action Games} \label{sec_Three}
Next, we consider $m$-3MA with $m>2$, where the learning dynamics are more complicated than the case of $m=2$. Now, Fig.~\ref{F04} shows that these dynamics are classified by not only $\alpha$ but $\gamma$.


\paragraph{Conservation of action number ($\gamma=0$)}: Let us consider the case of $\gamma=0$ (see the three panels in the middle row of Fig.~\ref{F04}). This case is viewed as the extension of the above dynamics to general $m>2$. In other words, when $\alpha=0$, the learning dynamics show a cycling behavior. When $\alpha>0$, players synchronize their actions throughout the learning: convergence to the pure-strategy equilibria, i.e., $\set_{\rm P}(m)$, depending on their initial strategies. When $\alpha<0$, players learn to desynchronize their actions and cannot reach the only uniform-choice equilibrium. These various dynamics are characterized by $V=\sum_{i}x_iy_iz_i-1/m^2$, again. We obtain the following theorem.

\begin{theorem}[Monotonicity of $V$ for $m$-3MA]
\label{thm_Monotonicity_m3MA}
In the replicator dynamics in Lem.~\ref{lem_RDGA}, except for the uniform-choice equilibrium $\bs{x}=\bs{y}=\bs{z}\in\set_{\rm U}(m)$, $\sign(\dot{V})=\sign(\alpha)$ holds.
\end{theorem}

See Appendix.~\ref{proof_thm_Monotonicity_m3MA} for its complete proof.

\textsc{Proof Sketch}. We calculate $\dot{V}$ by substituting Eq.~\eqref{ReplicatorDynamics}. Because $\gamma$ is ignored, only the terms of $\alpha$ and $\beta$ contribute to the dynamics. Here, the term of $\beta$ is negligible by the cyclic symmetry among the three players. Furthermore, the term of $\alpha$ is described as the variance of $y_iz_i$, which is positive everywhere except for $\bs{x}=\bs{y}=\bs{z}\in\set_{\rm U}(m)$. \qed

\paragraph{Convergence to two-action games ($\gamma>0$):} We next see the case of $\gamma>0$ (see the three panels in the upper row of Fig.~\ref{F04}). The dynamics approach those in the two-action games. As a simple example, see the dynamics in $\alpha=0$. Near the initial time, all players take all three actions stochastically. Throughout learning, however, one of the three actions ($a_2$ in the panel) becomes not played at all. Eventually, all players learn to use only the other two actions oscillatory. In other words, the three-action game converges to a two-action game after a sufficiently long time. The cases of $\alpha\neq 0$ can be interpreted similarly. In $\alpha>0$, the players first learn not to use one of three actions ($a_1$ in the panel) and synchronize their actions after that. In $\alpha<0$, players first learn not to use one action ($a_2$ in the panel) and finally reach a heteroclinic cycle.

\paragraph{Divergence to $m$-actions games ($\gamma<0$):} We finally see the case of $\gamma<0$ (see the three panels in the lower row of Fig.~\ref{F04}). The three players basically learn to desynchronize their actions. Indeed, the lower-middle and lower-right panels are similar to the middle-right panel ($\gamma=0, \alpha<0$). The case of $\gamma<0, \alpha>0$ is an exception. In this case, three players first desynchronize their actions (i.e., $V$ decreases in the lower-left panel), but synchronize their actions in the end (i.e., $V$ conversely increases). Because $\alpha(>0)$ and $\gamma(<0)$ conflict in this case, non-monotonic dynamics of $V$ are clearly observed. Whether their actions eventually synchronize or desynchronize depends on the initial condition.

\paragraph{$\gamma$ is the force to seek competition:} We now interpret another three-player interaction, $\gamma=a+b-2c$. First, note that in games with more than two actions, three players do not have to interact, i.e., they may choose different three actions separately. When two of the players compete with each other, i.e., choose the same action, the three players totally obtain the payoff of $a+b+c$. On the other hand, when they choose different three actions, their total payoff is $3c$. Thus, $\gamma>0\Leftrightarrow a+b+c>3c$ means that the players expectedly obtain higher payoffs when they compete. In this meaning, they, over time, learn to compete more often and to cyclically choose only two of $m$ actions. On the other hand, in $\gamma<0$, they learn to avoid competition as much as possible, leading to the dispersion of their action choices.

\section{Conclusion}
Three-player games are still much unexplored, compared to two-player games. This study focused on how three-player interactions affect the games. Surprisingly, we fully analyzed the Nash equilibria, even though solving the Nash equilibria of three-player games is sufficiently difficult in general~\cite{chen2006settling, daskalakis2009complexity}. We also found that the Nash equilibria are various and complex when three-player interactions, i.e., $\alpha$ and $\gamma$, conflict. Such three-player interactions also complicate the learning dynamics: The conserved quantities in a constant-sum game are no longer robust for such three-player interaction. We found the Lyapunov function $V$ and proved that $V$ monotonically increases or decreases depending on $\alpha$. We further demonstrated by simulation that learning dynamics can be classified by $\alpha$, the force to synchronize the choices of three players, and $\gamma$, the preference for competition with others. In conclusion, this study performed comprehensive theoretical and experimental analyses regarding three-player interactions for three-player matching $m$-action games.

How to achieve convergence to the Nash equilibrium, called last-iterate convergence, in three-player games is a topic of great interest. Recent literature shows that the FTRL achieves convergence to the Nash equilibrium in two-player zero-sum games when it optimistically foresees its opponent's future strategy~\cite{mertikopoulos2019optimistic, daskalakis2019last} or incorporates mutations into its learning~\cite{abe2022mutation, abe2023last}. In addition, memory, i.e., the ability to change one's action depending on the past games, is known to change game structures. For example, this memory extends the region of the Nash equilibrium~\cite{fujimoto2019emergence, fujimoto2021exploitation, usui2021symmetric, ueda2023memory}, induces divergence~\cite{fujimoto2023learning}, and achieves convergence~\cite{fujimoto2024memory}. As future work, it would be interesting to see how these learning algorithms perform in three-player games. This study will provide both a theoretical and experimental basis for such future works.

\begin{acks}
K. Ariu is supported by JSPS KAKENHI Grant No.~23K19986.
\end{acks}

\balance


\newpage

\appendix
\onecolumn

\begin{center}
{\LARGE\bf Appendix}
\end{center}

\section{Proofs}
\subsection{Proof of Theorem~\ref{thm_Nash}} \label{proof_thm_Nash}
\begin{proof}
First, by the following lemma (see Appendix.~\ref{proof_lem_Symmetry} for its proof), all the players take the same strategies in the Nash equilibria:
\begin{lemma}[Symmetry among players in the Nash equilibria]
\label{lem_Symmetry}
For any Nash equilibrium, $\bs{x}^*=\bs{y}^*=\bs{z}^*$ is satisfied.
\end{lemma}

Then, we show the second statement of the theorem.
From the definition, the Nash equilibria $(\bs{x}^*,\bs{y}^*,\bs{z}^*)$ should satisfy, with some constants $C_X, C_Y, C_Z\in\mathbb{R}$ for all $i\in\{1,\cdots,m\}$, the conditions of \eqref{Eq.-X}, \eqref{Eq.-Y}, and \eqref{Eq.-Z};
\begin{align}
    \begin{cases}
        x_i^*>0 &\Rightarrow f(y_i^*,z_i^*)=C_X\\
        x_i^*=0 &\Rightarrow f(y_i^*,z_i^*)\le C_X\\
    \end{cases},
    \tag{Eq.-X}\label{Eq.-X}\\
    \begin{cases}
        y_i^*>0 &\Rightarrow f(z_i^*,x_i^*)=C_Y\\
        y_i^*=0 &\Rightarrow f(z_i^*,x_i^*)\le C_Y\\
    \end{cases},
    \tag{Eq.-Y}\label{Eq.-Y}\\
    \begin{cases}
        z_i^*>0 &\Rightarrow f(x_i^*,y_i^*)=C_Z\\
        z_i^*=0 &\Rightarrow f(x_i^*,y_i^*)\le C_Z\\
    \end{cases}.
    \tag{Eq.-Z}\label{Eq.-Z}
\end{align}
By Lemma~\ref{lem_Symmetry}, $f(x_i^*,y_i^*)=f(y_i^*,z_i^*)=f(z_i^*,x_i^*)$ trivially holds. Thus, we newly define a function
\begin{align}
    \tilde{f}(x_i^*):=f(x_i^*,x_i^*)=(\alpha-\gamma)x_i^{*2}+\gamma x_i^*.
\end{align}
The Nash equilibrium conditions of~\eqref{Eq.-X}, \eqref{Eq.-Y}, and \eqref{Eq.-Z} are that there is $C_X=C_Y=C_Z=:C$ such that, with some constant $C\in\mathbb{R}$ for all $i\in\{1,\cdots,m\}$,
\begin{align}
    \begin{cases}
        x_i^*>0 &\Rightarrow \tilde{f}(x_i^*)=C \\
        x_i^*=0 &\Rightarrow \tilde{f}(x_i^*)\le C \\
    \end{cases}.
\end{align}
This condition is solved by the following lemmas (see Appendix~\ref{proof_lem_Interior} and \ref{proof_lem_Boundary} for their proofs). Combining these lemmas, we have proved Thm.~\ref{thm_Nash}

\begin{lemma}[The interior Nash equilibria]
\label{lem_Interior}
In the interior of the strategy spaces, i.e, $\bs{x},\bs{y},\bs{z}\in\interior(\Delta^{m-1})$, the set of $\bs{x}^*$ is given by
\begin{align}
    \begin{cases}
        \alpha=\gamma=0 &\Rightarrow \interior (\Delta^{m-1}) \\
        0\ge\alpha, 0\ge\gamma &\Rightarrow \set_{\rm U}(m) \\
        \alpha>0>\gamma &\Rightarrow \set_{\rm U}(m)\cup\set_{\rm DR}(m) \\
    \end{cases}
\end{align}
\end{lemma}

\begin{lemma}[The boundary Nash equilibria]
\label{lem_Boundary}
On the boundary of the strategy spaces, i.e, $\bs{x},\bs{y},\bs{z}\in\partial\Delta^{m-1}$, the set of $\bs{x}^*$ is given by
\begin{align}
    \begin{cases}
        \alpha=\gamma=0 &\Rightarrow \partial \Delta^{m-1}\\
        0\ge\alpha, 0\ge\gamma &\Rightarrow \emptyset \\
        \alpha\ge 0, \gamma\ge 0 &\Rightarrow \set_{\rm P}(m)\cup(\cup_{m'\in A}\projection(\set_{\rm U}(m'))) \\
        \alpha>0>\gamma &\Rightarrow \set_{\rm P}(m)\cup(\cup_{m'\in A^{(-)}}\projection(\set_{\rm U}(m'))) \\
        \gamma>0>\alpha &\Rightarrow (\cup_{m'\in A^{(+)}}\projection(\set_{\rm U}(m')))\cup(\cup_{m'\in A}\projection(\set_{\rm DR}(m'))) \\
    \end{cases}.
\end{align}
\end{lemma}
\end{proof}

\subsection{Proof of Lemma~\ref{lem_Symmetry}} \label{proof_lem_Symmetry}
\begin{proof}
We prove $\bs{x}^*=\bs{y}^*=\bs{z}^*$ by a contradiction method. First, we derive a contradiction from $\bs{x}^*\neq\bs{y}^*\neq\bs{z}^*\neq\bs{x}^*\Leftrightarrow (\bs{x}^*\neq\bs{y}^*)\land(\bs{y}^*\neq\bs{z}^*)\land(\bs{z}^*\neq\bs{x}^*)$. Second, we derive a contradiction from $\bs{x}^*=\bs{y}^*\neq\bs{z}^*$, which is equivalent to $\bs{y}^*=\bs{z}^*\neq\bs{x}^*$ and $\bs{z}^*=\bs{x}^*\neq\bs{y}^*$ by the cyclic symmetry of the three players. Thus, we can derive $\bs{x}^*=\bs{y}^*=\bs{z}^*$. Before such a contradiction method, we make some preparations for it.

\paragraph{Preparation for contradiction method:}
$f(y_i^*,z_i^*)$ is written as
\begin{align}
    f(y_i^*,z_i^*)=k_0y_i^*z_i^*+k_{+}y_i^*+k_{-}z_i^*,
\end{align}
where we defined $k_0:=\alpha-\gamma$, $k_{+}:=a-c=(\beta+\gamma)/2$, and $k_{-}:=b-c=(-\beta+\gamma)/2$. Here, remember $b<c<a$ and $b<\epsilon<a$. Thus, $k_{+}>0$, $k_{-}<0$, and $k_{+}>k_0>k_{-}$ hold. In the following, we prove the following four conditions and extend these conditions;
\begin{align}
    (x_i^*\ge y_i^*>z_i^*>0) \lor (x_i^*>y_i^*\ge z_i^*>0) &\Rightarrow (C_Z>C_Y)\land(C_X>C_Y),
    \tag{A1}\label{A1_origin} \\
    x_i^*\ge y_i^*>z_i^*=0 &\Rightarrow (C_Z>C_Y)\land(C_X>C_Y),
    \tag{B1}\label{B1_origin} \\
    y_i^*\ge x_i^*>z_i^*=0 &\Rightarrow C_X>C_Y,
    \tag{B6}\label{B6_origin} \\
    x_i^*>y_i^*=z_i^*=0 &\Rightarrow C_Z>C_X.
    \tag{C1}\label{C1_origin}
\end{align}

\paragraph{Proof of~\eqref{A1_origin}:}
Since $x_i^*\neq 0$, $y_i^*\neq 0$, and $z_i^*\neq 0$, the Nash equilibrium condition is
\begin{align}
    \begin{cases}
        C_X=f(y_i^*,z_i^*)=k_0y_i^*z_i^*+k_{+}y_i^*+k_{-}z_i^*\\
        C_Y=f(z_i^*,x_i^*)=k_0z_i^*x_i^*+k_{+}z_i^*+k_{-}x_i^*\\
        C_Z=f(x_i^*,y_i^*)=k_0x_i^*y_i^*+k_{+}x_i^*+k_{-}y_i^*\\
    \end{cases}.
    \label{Nash_xyz_nonzero}
\end{align}
Then, $C_Z>C_Y$ is because
\begin{align}
    \begin{cases}
        k_0\ge 0 &\Rightarrow C_Z=k_0x_i^*\underbrace{y_i^*}_{\ge z_i^*}+k_{+}\underbrace{x_i^*}_{>z_i^*}+\underbrace{k_{-}y_i^*}_{\ge k_{-}x_i^*}>k_0z_i^*x_i^*+k_{+}z_i^*+k_{-}x_i^*=C_Y \\
        k_0\le 0 &\Rightarrow C_Z=\underbrace{(k_0y_i^*+k_{+})}_{\ge (k_0x_i^*+k_{+})}\underbrace{x_i^*}_{>z_i^*}+\underbrace{k_{-}y_i^*}_{\ge k_{-}x_i^*}>(k_0x_i^*+k_{+})z_i^*+k_{-}x_i^*=C_Y \\
    \end{cases}
    \label{A1_proof1}
\end{align}
Here, we used $k_0y_i^*+k_{+}\ge k_0x_i^*+k_{+}\ge k_0+k_{+}>0$ for $k_0\le 0$. On the other hand, $C_X>C_Y$ is because
\begin{align}
    \begin{cases}
        k_0\ge 0 &\Rightarrow C_X=\underbrace{(k_0y_i^*+k_{-})z_i^*}_{>(k_0z_i^*+k_{-})x_i^*}+k_{+}\underbrace{y_i^*}_{\ge z_i^*}>(k_0z_i^*+k_{-})x_i^*+k_{+}z_i^*=C_Y \\
        k_0\le 0 &\Rightarrow C_X=\underbrace{k_0y_i^*}_{\ge k_0x_i^*}z_i^*+k_{+}\underbrace{y_i^*}_{\ge z_i^*}+\underbrace{k_{-}z_i^*}_{>k_{-}x_i^*}>k_0z_i^*x_i^*+k_{+}z_i^*+k_{-}x_i^*=C_Y \\
    \end{cases}
    \label{A1_proof2}
\end{align}
Here, $(k_0y_i^*+k_{-})z_i^*>(k_0z_i^*+k_{-})x_i^*$ is because $k_0z_i^*+k_{-}\le k_0y_i^*+k_{-}\le k_0+k_{-}<0$ for $k_0\ge 0$ and $z_i^*<x_i^*$.

\paragraph{Proof of~\eqref{B1_origin}:}
Since $x_i^*\neq 0$, $y_i^*\neq 0$, and $z_i^*=0$, the Nash equilibrium condition is
\begin{align}
    \begin{cases}
        C_X=f(y_i^*,z_i^*)=k_{+}y_i^*\\
        C_Y=f(z_i^*,x_i^*)=k_{-}x_i^*\\
        C_Z\ge f(x_i^*,y_i^*)=k_0x_i^*y_i^*+k_{+}x_i^*+k_{-}y_i^*\\
    \end{cases}.
    \label{Nash_xy_nonzero}
\end{align}
Here, $C_X>C_Y$ trivially holds, while $C_Z>C_Y$ is because
\begin{align}
    C_Z\ge \underbrace{(k_0y_i^*+k_{+})}_{>0}x_i^*+k_{-}\underbrace{y_i^*}_{>x_i^*}>k_{-}x_i^*=C_Y.
\end{align}

\paragraph{Proof of~\eqref{B6_origin}:}
The Nash equilibrium condition is given by Eqs.~\eqref{Nash_xyz_nonzero}, again, and $C_X>C_Y$ trivially holds.

\paragraph{Proof of~\eqref{C1_origin}:}
Since $x_i^*\neq 0$, $y_i^*=z_i^*=0$, the Nash equilibrium condition is
\begin{align}
    \begin{cases}
        C_X=f(y_i^*,z_i^*)=0\\
        C_Y\ge f(z_i^*,x_i^*)=k_{-}x_i^*\\
        C_Z\ge f(x_i^*,y_i^*)=k_{+}x_i^*\\
    \end{cases}.
    \label{Nash_x_nonzero}
\end{align}
Here, $C_Z>C_X$ trivially holds.

\paragraph{Extension of all the obtained conditions:}
These conditions of Eqs.~\eqref{A1_origin}, \eqref{B1_origin}, \eqref{B6_origin}, and \eqref{C1_origin} are extended as follows;
\begin{align}
    (x_i^*\ge y_i^*>z_i^*>0) \lor (x_i^*>y_i^*\ge z_i^*>0) &\Rightarrow (C_Z>C_Y)\land(C_X>C_Y), 
    \tag{A1}\label{A1}\\
    (y_i^*\ge z_i^*>x_i^*>0) \lor (y_i^*>z_i^*\ge x_i^*>0) &\Rightarrow (C_X>C_Z)\land(C_Y>C_Z), 
    \tag{A2}\label{A2}\\
    (z_i^*\ge x_i^*>y_i^*>0) \lor (z_i^*>x_i^*\ge y_i^*>0) &\Rightarrow (C_Y>C_X)\land(C_Z>C_X), 
    \tag{A3}\label{A3}\\
    (z_i^*\ge y_i^*>x_i^*>0) \lor (z_i^*>y_i^*\ge x_i^*>0) &\Rightarrow (C_Y>C_Z)\land(C_Y>C_X), 
    \tag{A4}\label{A4}\\
    (x_i^*\ge z_i^*>y_i^*>0) \lor (x_i^*>z_i^*\ge y_i^*>0) &\Rightarrow (C_Z>C_X)\land(C_Z>C_Y), 
    \tag{A5}\label{A5}\\
    (y_i^*\ge x_i^*>z_i^*>0) \lor (y_i^*>x_i^*\ge z_i^*>0) &\Rightarrow (C_X>C_Y)\land(C_X>C_Z), 
    \tag{A6}\label{A6}\\
    x_i^*\ge y_i^*>z_i^*=0 &\Rightarrow (C_Z>C_Y)\land(C_X>C_Y), 
    \tag{B1}\label{B1}\\
    y_i^*\ge z_i^*>x_i^*=0 &\Rightarrow (C_X>C_Z)\land(C_Y>C_Z), 
    \tag{B2}\label{B2}\\
    z_i^*\ge x_i^*>y_i^*=0 &\Rightarrow (C_Y>C_X)\land(C_Z>C_X), 
    \tag{B3}\label{B3}\\
    z_i^*\ge y_i^*>x_i^*=0 &\Rightarrow C_Y>C_Z, 
    \tag{B4}\label{B4}\\
    x_i^*\ge z_i^*>y_i^*=0 &\Rightarrow C_Z>C_X, 
    \tag{B5}\label{B5}\\
    y_i^*\ge x_i^*>z_i^*=0 &\Rightarrow C_X>C_Y, 
    \tag{B6}\label{B6}\\
    x_i^*>y_i^*=z_i^*=0 &\Rightarrow C_Z>C_X, 
    \tag{C1}\label{C1}\\
    y_i^*>z_i^*=x_i^*=0 &\Rightarrow C_X>C_Y, 
    \tag{C2}\label{C2}\\
    z_i^*>x_i^*=y_i^*=0 &\Rightarrow C_Y>C_Z.
    \tag{C3}\label{C3}
\end{align}
These conditions are obtained by using symmetries. First, the condition of~\eqref{A4} is obtained by reversing all the equal signs in Eqs.~\eqref{A1_proof1} and \eqref{A1_proof2}. Let $\sigma_{\rm c}$ denote the cyclic permutation for all the parameters of X, Y, and Z. Then, we obtain all the other conditions by $\eqref{A2}=\sigma_{\rm c}(\eqref{A1})$, $\eqref{A3}=\sigma_{\rm c}^2(\eqref{A1})$, $\eqref{A5}=\sigma_{\rm c}(\eqref{A4})$, $\eqref{A6}=\sigma_{\rm c}^2(\eqref{A4})$, $\eqref{B2}=\sigma_{\rm c}(\eqref{B1})$, $\eqref{B3}=\sigma_{\rm c}^2(\eqref{B1})$, $\eqref{B4}=\sigma_{\rm c}(\eqref{B6})$, $\eqref{B5}=\sigma_{\rm c}^2(\eqref{B6})$, $\eqref{C2}=\sigma_{\rm c}(\eqref{C1})$, and $\eqref{C3}=\sigma_{\rm c}^2(\eqref{C1})$.

\paragraph{Contradiction of $\bs{x}^*\neq \bs{y}^*\neq \bs{z}^*\neq \bs{x}^*$:}
Let us assume $\bs{x}^*\neq \bs{y}^*\neq \bs{z}^*\neq \bs{x}^*\Leftrightarrow (\bs{x}^*\neq \bs{y}^*)\land(\bs{y}^*\neq \bs{z}^*)\land(\bs{z}^*\neq \bs{x}^*)$. If so, there are $i_1$, $i_2$, $i_3\in\{1,\cdots,m\}$ such that $(x_{i_1}^*>y_{i_1}^*) \land (y_{i_2}^*>z_{i_2}^*) \land (z_{i_3}^*>x_{i_3}^*)$. Here, we obtain
\begin{align}
    x_{i_1}^*>y_{i_1}^* &\Rightarrow \eqref{A1}\lor \eqref{A3}\lor \eqref{A5}\lor \eqref{B1}\lor \eqref{B3}\lor \eqref{B5}\lor \eqref{C1}
    \nonumber \\
    &\Rightarrow (C_Z>C_X) \lor ((C_Z>C_Y)\land(C_X>C_Y))
    \nonumber \\
    &\Rightarrow (C_Z>C_X) \lor (C_Z>C_Y),
    \label{con_x>y_positive} \\
    y_{i_2}^*>z_{i_2}^* &\Rightarrow \eqref{A2}\lor \eqref{A1}\lor \eqref{A6}\lor \eqref{B2}\lor \eqref{B1}\lor \eqref{B6}\lor \eqref{C2}
    \nonumber \\
    &\Rightarrow (C_X>C_Y) \lor ((C_X>C_Z)\land(C_Y>C_Z))
    \nonumber \\
    &\Rightarrow (C_X>C_Y) \lor (C_X>C_Z),
    \label{con_y>z_positive} \\
    z_{i_3}^*>x_{i_3}^* &\Rightarrow \eqref{A3}\lor \eqref{A2}\lor \eqref{A4}\lor \eqref{B3}\lor \eqref{B2}\lor \eqref{B4}\lor \eqref{C3}
    \nonumber \\
    &\Rightarrow (C_Y>C_Z) \lor ((C_Y>C_X)\land(C_Z>C_X))
    \nonumber \\
    &\Rightarrow (C_Y>C_Z) \lor (C_Y>C_X).
    \label{con_z>x_positive}
\end{align}
Here, the conditions of~\eqref{con_x>y_positive}, \eqref{con_y>z_positive}, and \eqref{con_z>x_positive} cannot be satisfied simultaneously. Thus, such $i_1$, $i_2$, and $i_3$ do not exist. This is a contradiction, proving that $\bs{x}^*\neq \bs{y}^*\neq \bs{z}^*\neq \bs{x}^*$ cannot hold.

\paragraph{Contradiction of $\bs{x}^*=\bs{y}^*\neq \bs{z}^*$:}
Second, we assume $\bs{x}^*=\bs{y}^*\neq \bs{z}^*$. If so, there are $i_1$, $i_2\in\{1,\cdots,m\}$ such that $(x_{i_1}=y_{i_1}>z_{i_1}) \land (z_{i_2}>x_{i_2}=y_{i_2})$. Here, we obtain
\begin{align}
    x_{i_1}^*=y_{i_1}^*>z_{i_1}^* &\Rightarrow (\eqref{A1}\land \eqref{A6})\lor (\eqref{B1}\land \eqref{B6})
    \nonumber \\
    &\Rightarrow (C_Z>C_Y)\land(C_X>C_Y)
    \nonumber \\
    &\Rightarrow C_Z>C_Y,
    \label{con_x=y>z_positive} \\
    z_{i_2}^*>x_{i_2}^*=y_{i_2}^* &\Rightarrow (\eqref{A3}\land \eqref{A4})\lor \eqref{C3}
    \nonumber \\
    &\Rightarrow C_Y>C_Z.
    \label{con_z>x=y_positive}
\end{align}
Here, the conditions~\eqref{con_x=y>z_positive} and \eqref{con_z>x=y_positive} cannot be satisfied simultaneously. Thus, such $i_1$ and $i_2$ do not exist. This is a contradiction, proving that $\bs{x}^*=\bs{y}^*\neq \bs{z}^*$ cannot hold. Finally, because we proved that neither $\bs{x}^*\neq \bs{y}^*\neq \bs{z}^*\neq \bs{x}^*$ nor $\bs{x}^*=\bs{y}^*\neq \bs{z}^*$ hold, we prove $\bs{x}^*=\bs{y}^*=\bs{z}^*$.
\end{proof}

\subsection{Proof of Lemma~\ref{lem_Interior}} \label{proof_lem_Interior}
\begin{proof}
In the interior of the strategy space, the condition for $\bs{x}^*$ is, with some $C$ for all $i\in\{1,\cdots,m\}$,
\begin{align}
    \tilde{f}(x_i^*)=C.
\end{align}

\paragraph{Classification depending on parameters:} The Nash equilibrium condition depends on the function of $\tilde{f}$. Fig.~\ref{FS01} classifies the function by $\alpha$ and $\gamma$. We define the cases from 1) to 9) as follows.
\begin{align}
    1).&\ \alpha>\gamma\ge 0,\quad 2).\ \alpha>0>\gamma,\quad 3).\ 0\ge\alpha>\gamma, \\
    4).&\ \alpha=\gamma> 0,\quad 5).\ \alpha=\gamma=0,\quad 6).\ 0>\alpha=\gamma, \\
    7).&\ \gamma>\alpha\ge 0,\quad 8).\ \gamma>0>\alpha,\quad 9).\ 0\ge\gamma>\alpha.
\end{align}

\begin{figure}[h!]
    \centering
    \includegraphics[width=0.7\hsize]{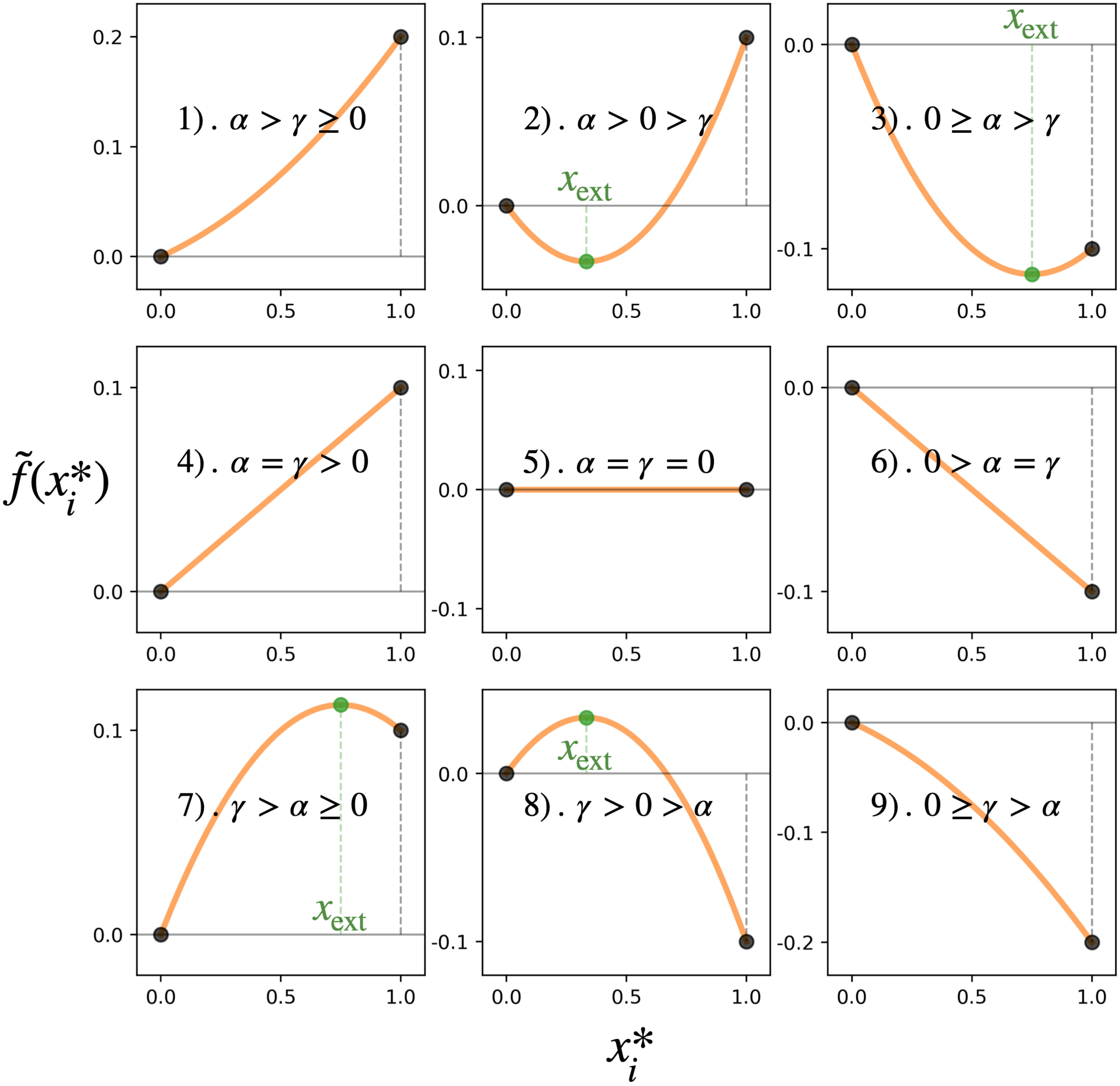}
    \caption{All possible $\tilde{f}(x_i^*)$ depending on $\alpha$ and $\gamma$. The orange lines show $\tilde{f}(x_i^*)$ for the horizontal axis of $x_i^*$. The black dots are the values in $x_i^{*}=0$ and $1$. The green dots show the extreme value in $x_{\ext}:=\gamma/\{2(\gamma-\alpha)\}$; $0<x_{\ext}<1/2$ in the cases of 2) and 8), while $1/2<x_{\ext}<1$ in the cases of 3) and 7).}
    \label{FS01}
\end{figure}

\paragraph{Case 5).} In this case, $\tilde{f}(x_i^*)=0$ always holds. Thus, all the interior points of the strategy space are the Nash equilibrium, i.e., $\bs{x}^*\in\interior(\Delta^{m-1})$.

\paragraph{Case 1) 3) 4) 6) 7) 9).} In these cases, $\tilde{f}(x_i^*)=C$ has no multiple solutions in $0<x_i^*<1/2$. Thus, the interior Nash equilibrium is only $\bs{x}^*\in\{\bs{1}/m\}=:\set_{\rm U}(m)$ (U: Uniform-choice equilibrium).

\paragraph{Case 2) 8).} In these cases, $\tilde{f}(x_i^*)$ is parabolic and takes its extreme value at $x_i^*=\gamma/\{2(\gamma-\alpha)\}=:x_{\ext}$, which is in the region of $0<x_{\ext}<1/2$. Thus, for some $C$, $\tilde{f}(x_i^*)=C$ has double roots $x_i^*=x_{+},x_{-}$, described as
\begin{align}
    x_{+}=x_{\ext}+\delta,\ x_{-}=x_{\ext}-\delta,
\end{align}
with $0<\delta<x_{\ext}$. Here, a condition for a strategy which composes of $k$-pieces of $x_{+}$ and $(m-k)$-pieces of $x_{-}$ for $k\in\{1,\cdots,m-1\}$, i.e.,
\begin{align}
    \bs{x}_{{\rm DR};k}:=(\underbrace{x_{+},\cdots,x_{+}}_{k},\underbrace{x_{-},\cdots,x_{-}}_{m-k})
\end{align}
to be the Nash equilibrium is
\begin{align}
    kx_{+}+(m-k)x_{-}=1 &\Leftrightarrow mx_{\ext}-(m-2k)\delta=1 \\
    &\Leftrightarrow \delta=\frac{mx_{\ext}-1}{m-2k}.
\end{align}
Since $0<\delta\le x_{\ext}$ should be satisfied, we solve
\begin{align}
    0<\frac{mx_{\ext}-1}{m-2k}<x_{\ext}\Leftrightarrow \begin{cases}
        \displaystyle k<\frac{1}{2x_{\ext}} & \displaystyle \left(x_{\ext}>\frac{1}{m}\right) \\
        \displaystyle k>\frac{1}{2x_{\ext}} & \displaystyle \left(x_{\ext}<\frac{1}{m}\right) \\
    \end{cases}.
\end{align}
In conclusion, the interior Nash equilibria include
\begin{align}
    \bs{x}^*&\in\begin{cases}
        \{\sigma(\bs{x}_{{\rm DR};k})|\sigma\in S_m, k=1,\cdots,\lfloor 1/(2x_{\ext})\rfloor\} & (x_{\ext}>1/m) \\
        \{\sigma(\bs{x}_{{\rm DR};k})|\sigma\in S_m, k=\lceil 1/(2x_{\ext})\rceil,\cdots,m-1\} & (x_{\ext}<1/m) \\
    \end{cases}\\
    &=:\set_{\rm DR}(m),
\end{align}
(DR: Double Roots equilibria). Here, $\sigma$ is a permutation function in the $m$-dimensional symmetric group $S_m$. Since the uniform-choice equilibrium is also included, all the interior Nash equilibria are given by $\set_{\rm U}(m)\cup\set_{\rm DR}(m)$.
\end{proof}

\subsection{Proof of Lemma~\ref{lem_Boundary}} \label{proof_lem_Boundary}
\begin{proof}
In the interior of the strategy space, the condition for $\bs{x}^*$ is, with some $C$ for all $i\in\{1,\cdots,m\}$,
\begin{align}
    \begin{cases}
        x_i^*>0 &\Rightarrow \tilde{f}(x_i^*)=C \\
        x_i^*=0 &\Rightarrow \tilde{f}(x_i^*)\le C \\
    \end{cases}.
\end{align}
This is equivalent to, with some $C\ge 0 (\Leftrightarrow \tilde{f}(0)\le C)$ for all $i$ such that $x_i^*>0$,
\begin{align}
    \tilde{f}(x_i^*)=C.
\end{align}
In the following, we consider this condition under the classification of $\alpha$ and $\gamma$ (see Fig.~\ref{FS01} again).

\paragraph{Case 5).} In this case, $\tilde{f}(x_i^*)=0$ always holds. Thus, all the boundary points of the strategy space are the Nash equilibrium, i.e., $\bs{x}^*\in\partial\Delta^{m-1}$.

\paragraph{Case 3) 6) 9).} In these cases, $\tilde{f}(x_i^*)<0=\tilde{f}(0)$ always holds for $0<x_i^*\le 1$. Thus, there is no boundary Nash equilibrium.

\paragraph{Case 1) 4) 7).} In these cases, $\tilde{f}(x_i^*)\ge 0=\tilde{f}(0)$ is positive for all $0\le x_i^*\le 1$. In other words, $\tilde{f}(1)\ge \tilde{f}(0)$ holds, meaning that the boundary Nash equilibria include $\{\bs{e}_1,\cdots,\bs{e}_m\}=:\set_{\rm P}(m)$ (P: Pure-strategy equilibria). Furthermore, for $2\le m'\le m-1$, $\tilde{f}(1/m')\ge \tilde{f}(0)$ also holds so that the boundary Nash equilibria also include $\projection(\set_{\rm U}(m'))$ for all $m'\in\{2,\cdots,m-1\}=:A$. Here, $\projection$ shows the inverse projection from $m'<m$- to $m$-dimensional space with arbitrary permutations. Thus, the Nash equilibria are given by
\begin{align}
    \set_{\rm P}(m)\cup\left(\bigcup_{m'\in A}\projection(\set_{\rm U}(m'))\right).
\end{align}

\paragraph{case 2).} In this case, $\tilde{f}(x_i^*)\ge 0=\tilde{f}(0)$ holds for $x_{\ext}\le x_i^*\le 1$. Thus, since $\tilde{f}(1)\ge \tilde{f}(0)$ holds, the Nash equilibria include $\set_{\rm P}(m)$. Furthermore, since $\tilde{f}(1/m')\ge \tilde{f}(0)$ holds for $2\le m'\le 1/x_{\ext}$, the Nash equilibria include $\projection(\set_{\rm U}(m'))$ for all $m'\in\{2,\cdots,\min(\lfloor 1/x_{\ext}\rfloor,m-1)\}=:A^{(-)}$. To summarize, the Nash equilibria are given by
\begin{align}
    \set_{\rm P}(m)\cup \left(\bigcup_{m'\in A^{(-)}}\projection(\set_{\rm U}(m'))\right).
\end{align}

\paragraph{Case 8).} In this case, $\tilde{f}(x_i^*)\ge 0=\tilde{f}(0)$ holds for $0<x_i^*\le x_{\ext}$. Thus, the Nash equilibria include $\projection(\set_{\rm DR}(m'))$ for all $m'\in A$. Furthermore, the Nash equilibria also include $\projection(\set_{\rm U}(m'))$ for all $m'\in\{\lceil 1/(2x_{\ext})\rceil,\cdots,m-1\}=:A^{(+)}$. To summarize, the Nash equilibria are given by $\set_{\rm U}(1/(2x_{\ext})\le m'\le m-1)\cup \set_{\rm DR}(2\le m'\le m-1)$.
\begin{align}
    \left(\bigcup_{m'\in A^{(+)}}\projection(\set_{\rm U}(m'))\right)\cup\left(\bigcup_{m'\in A}\projection(\set_{\rm DR}(m'))\right)
\end{align}
\end{proof}

\subsection{Proof of Lemma~\ref{lem_Dynamics_two}} \label{proof_lem_Dynamics}
\begin{proof}
Two-action games give $\bs{x}=(x_1,x_2)$, so that $x_2=\bar{x}_1$ holds. Similar equations hold for Y and Z. First, Eq.~\eqref{FTRL3} in the FTRL algorithm is calculated as
\begin{align}
    q_1(\bs{x}^{\dagger})&=\arg\max_{\bs{x}}\{\bs{x}^{\dagger}\cdot\bs{x}-h(\bs{x})\}
    \nonumber \\
    &=\arg\max_{x_1}\{x_1^{\dagger}x_1+x_2^{\dagger}(1-x_1)-h_{\two}(x_1)\}
    \nonumber \\
    &=\arg\max_{x_1}\{\underbrace{(x_1^{\dagger}-x_2^{\dagger})}_{=:\Delta x^{\dagger}}x_1-h_{\two}(x_1)\}
    \label{q_FTRL_two}
\end{align}
Thus, $q_1(\bs{x}^{\dagger})$ is determined only by the single variable of $\Delta x^{\dagger}$, which dynamics are
\begin{align}
    \Delta\dot{x}^{\dagger}&=\dot{x}_1^{\dagger}-\dot{x}_2^{\dagger}
    \nonumber \\
    &=\frac{\partial u}{\partial x_1}-\frac{\partial u}{\partial x_2}
    \nonumber \\
    &=f(y_1,z_1)-f(y_2,z_2)
    \nonumber \\
    &=f(y_1,z_1)-f(1-y_1,1-z_1)
    \nonumber \\
    &=(\alpha+\beta)y_1+(\alpha-\beta)z_1-\alpha=:f_{\two}(y_1,z_1),
\end{align}
independent of $\gamma$.

We further assume $\bs{x}\in\interior\Delta^{m-1}$, i.e., $0<x_1<1$. Then, the extreme condition in Eq.~\eqref{q_FTRL_two} is calculated as
\begin{align}
    h_{\two}'(q_1(\bs{x}^{\dagger}))=\Delta x^{\dagger}\Leftrightarrow q_1(\bs{x}^{\dagger})=h_{\two}'^{-1}(\Delta x^{\dagger}).
\end{align}
By this equation, we obtain
\begin{align}
    \dot{x}_1&=\frac{\partial u}{\partial x_1}\frac{\partial q_1}{\partial x_1^{\dagger}}+\frac{\partial u}{\partial x_2}\frac{\partial q_1}{\partial x_2^{\dagger}}
    \nonumber \\
    &=\frac{\partial u}{\partial x_1}(h_{\two}'^{-1})'(\Delta x^{\dagger})-\frac{\partial u}{\partial x_2}(h_{\two}'^{-1})'(\Delta x^{\dagger})
    \nonumber \\
    &=(f(y_1,z_1)-f(y_2,z_2))(h_{\two}'^{-1})'(\Delta x^{\dagger})
    \nonumber \\
    &=f_{\two}(y_1,z_1)(h_{\two}'^{-1})'(\Delta x^{\dagger}),
\end{align}

Let us calculate $(h_{\two}'^{-1})'(\Delta x^{\dagger})$ as
\begin{align}
    &h_{\two}(x_1)=x_1\log x_1+(1-x_1)\log(1-x_1)
    \nonumber \\
    \Leftrightarrow &h_{\two}'(x_1)=\log x_1-\log (1-x_1)\ (=\Delta x^{\dagger})
    \nonumber \\
    \Leftrightarrow &h_{\two}'^{-1}(\Delta x^{\dagger})=\frac{\exp(\Delta x^{\dagger})}{1+\exp(\Delta x^{\dagger})}\ (=x_1)
    \nonumber \\
    \Leftrightarrow &(h_{\two}'^{-1})'(\Delta x^{\dagger})=\frac{\exp(\Delta x^{\dagger})}{1+\exp(\Delta x^{\dagger})}\frac{1}{1+\exp(\Delta x^{\dagger})}=x_1(1-x_1),
\end{align}
for the entropic regularizer $h(\bs{x})=\bs{x}\cdot\log\bs{x}$ and
\begin{align}
    &h_{\two}(x_1)=\frac{x_1^2+(1-x_1)^2}{2}
    \nonumber \\
    \Leftrightarrow &h_{\two}'(x_1)=2x_1-1\ (=\Delta x^{\dagger})
    \nonumber \\
    \Leftrightarrow &h_{\two}'^{-1}(\Delta x^{\dagger})=\frac{\Delta x^{\dagger}+1}{2}\ (=x_1)
    \nonumber \\
    \Leftrightarrow &(h_{\two}'^{-1})'(\Delta x^{\dagger})=\frac{1}{2},
\end{align}
for the Euclidean regularizer $h(\bs{x})=\|\bs{x}\|^2/2$. To summarize, we denote $(h_{\two}'^{-1})'(\Delta x^{\dagger})$ by
\begin{align}
    w(x_1)&:=\begin{cases}
        x_1(1-x_1) & (h(\bs{x})=\bs{x}\cdot\log\bs{x}) \\
        1/2 & (h(\bs{x})=\|\bs{x}\|^2/2) \\
    \end{cases}.
\end{align}
\end{proof}

\subsection{Proof of Theorem~\ref{thm_Monotonicity}} \label{proof_thm_Monotonicity}
\begin{proof}
In two-action games, we obtain
\begin{align}
    V(\bs{x},\bs{y},\bs{z})=x_1y_1z_1+(1-x_1)(1-y_1)(1-z_1).
\end{align}
We now define the cyclic sum $\cycle$ as
\begin{align}
    \cycle \mc{F}(\bs{x},\bs{y},\bs{z}):=\mc{F}(\bs{x},\bs{y},\bs{z})+\mc{F}(\bs{y},\bs{z},\bs{x})+\mc{F}(\bs{z},\bs{x},\bs{y}),
\end{align}
for arbitrary function $\mc{F}$. Using this $\cycle$, $V$ is calculated as
\begin{align}
    \dot{V}&=\cycle\dot{x}_1y_1z_1-\dot{x}_1(1-y_1)(1-z_1)
    \nonumber \\
    &=\cycle w(x_1)\{\alpha(y_1+z_1-1)+\beta(y_1-z_1)\}(y_1+z_1-1)
    \nonumber \\
    &=\alpha\cycle w(x_1)(y_1+z_1-1)^2+\beta w(x_1)(y_1-z_1)(y_1+z_1-1)
    \nonumber \\
    &=\alpha\cycle w(x_1)(y_1+z_1-1)^2,
\end{align}
Here, because $w(x_1)>0$ always holds, $\sign(\dot{V})=\sign(\alpha)$ is proved. In the last line of this equation, we used
\begin{align}
    &\cycle w(x_1)(y_1-z_1)(y_1+z_1-1)
    \nonumber \\
    &=\cycle x_1(1-x_1)(y_1-z_1)(y_1+z_1-1)
    \nonumber \\
    &=-\underbrace{\cycle x_1(y_1-z_1)}_{=0}-\underbrace{\cycle x_1^2(y_1^2-z_1^2)}_{=0}+\underbrace{\cycle (x_1y_1^2-z_1x_1^2)}_{=0}+\underbrace{\cycle (x_1^2y_1-z_1^2x_1)}_{=0}
    \nonumber \\
    &=0,
\end{align}
for the entropic regularizer $w(x_1)=x_1(1-x_1)$ and
\begin{align}
    &\cycle w(x_1)(y_1-z_1)(y_1+z_1-1)
    \nonumber \\
    &=\frac{1}{2}\cycle (y_1-z_1)(y_1+z_1-1)
    \nonumber \\
    &=\frac{1}{2}\underbrace{\cycle (y_1^2-z_1^2)}_{=0}-\frac{1}{2}\underbrace{\cycle (y_1-z_1)}_{=0}
    \nonumber \\
    &=0,
\end{align}
for the Euclidean regularizer $w(x_1)=1/2$.
\end{proof}

\subsection{Proof of Theorem~\ref{thm_Connection}} \label{proof_thm_Connection}
\begin{proof}
First, we can derive
\begin{align}
    \dot{G}&=\cycle\sumi \frac{\partial u}{\partial x_i}(x_i-x_i^*)
    \nonumber \\
    &=\cycle\sumi f(y_i,z_i)(x_i-x_i^*)
    \nonumber \\
    &=\cycle f_{\two}(y_1,z_1)(x_1-x_1^*)
    \nonumber \\
    &=\cycle \{\alpha(y_1+z_1-1)+\beta(y_1-z_1)\}(x_1-x_1^*)
    \nonumber \\
    &=\alpha\cycle(y_1+z_1-1)(x_1-x_1^*)
    \nonumber \\
    &=2\alpha \left\{(x_1y_1+y_1z_1+z_1x_1)-(x_1+y_1+z_1)+\frac{3}{4}\right\}
    \nonumber \\
    &=2\alpha V.
\end{align}
Here, in the fifth equal sign, the term of $\beta$ disappears because
\begin{align}
    &\cycle (y_1-z_1)(x_1-x_1^*)
    \nonumber \\
    &=\cycle (y_1-z_1)\left(x_1-\frac{1}{2}\right)
    \nonumber \\
    &=\underbrace{\cycle x_1(y_1-z_1)}_{=0} -\frac{1}{2}\underbrace{\cycle (y_1-z_1)}_{=0}
    \nonumber \\
    &=0.
\end{align}
The sixth equal sign holds because
\begin{align}
    V&=\sumi x_iy_iz_i-\frac{1}{m^2}
    \nonumber \\
    &=x_1y_1z_1+(1-x_1)(1-y_1)(1-z_1)-\frac{1}{4}
    \nonumber \\
    &=(x_1y_1+y_1z_1+z_1x_1)-(x_1+y_1+z_1)+\frac{3}{4}.
\end{align}
\end{proof}

\subsection{Proof of Theorem~\ref{thm_Monotonicity_m3MA}} \label{proof_thm_Monotonicity_m3MA}
\begin{proof}
\begin{align}
    \dot{V}&=\cycle \sumi \dot{x}_iy_iz_i
    \nonumber \\
    &=\cycle \sumi x_iy_iz_i(f(y_i,z_i)-\sumi x_if(y_i,z_i))
    \nonumber \\
    &=\alpha\cycle \sumi x_iy_iz_i(y_iz_i-\sumi x_iy_iz_i)+\frac{\beta}{2}\cycle \sumi x_iy_iz_i\{(y_i-z_i)-\sumi x_i(y_i-z_i)\}
    \nonumber \\
    &=\alpha\cycle \sumi x_iy_iz_i(y_iz_i-\sumi x_iy_iz_i)
    \nonumber \\
    &=\alpha\cycle \{\sumi x_i(y_iz_i)^2 - (\sumi x_iy_iz_i)^2\}
    \nonumber \\
    &=\alpha\cycle {\rm Var}[y_iz_i]_{x_i}.
\end{align}
Here, ${\rm Var}[y_iz_i]_{x_i}$ shows the variance of $y_iz_i$ based on the distribution of $x_i$ for $i$. In the fourth equal sign, we used
\begin{align}
    &\cycle \sumi x_iy_iz_i\{(y_i-z_i)-\sumi x_i(y_i-z_i)\}
    \nonumber \\
    &=\sumi x_iy_iz_i\{\underbrace{\cycle (y_i-z_i)}_{=0}-\sumi\underbrace{\cycle x_i(y_i-z_i)}_{=0}\}
    \nonumber \\
    &=0.
\end{align}

From the definition of variance, ${\rm Var}[y_iz_i]_{x_i}=0$ holds if and only if $y_iz_i$ takes a constant value $C_X$ for all $i$. Thus, the condition for $\cycle {\rm Var}[y_iz_i]_{x_i}=0$ is
\begin{align}
    \begin{cases}
        y_iz_i=C_X\\
        z_ix_i=C_Y\\
        x_iy_i=C_Z\\
    \end{cases},
\end{align}
for some constants $C_X$, $C_Y$, and $C_Z$. In the same way as the proof of Lem.~\ref{lem_Symmetry}, the condition is equivalent to $\bs{x}=\bs{y}=\bs{z}=\bs{1}/m$, meaning the uniform-choice equilibrium.
\end{proof}

\end{document}